\documentclass{osa-article}
\journal{oe}
\articletype{Research Article}
\usepackage{graphicx}
\usepackage{subcaption}
\usepackage{makecell}
\usepackage{mathtools}
\DeclareMathOperator*{\argmin}{argmin}
\usepackage{algorithm}
\usepackage{algpseudocode}
\usepackage{comment}
\begin{document}

\title{Undersampling Raster Scans in Spectromicroscopy for reduced dose and faster measurements}

\author{Oliver Townsend,\authormark{1,2,*} Silvia Gazzola,\authormark{1} Sergey Dolgov,\authormark{1} and Paul Quinn\authormark{2}}

\address{\authormark{1}Department of Mathematics, 4W, University of Bath, Claverton Down, Bath, BA2 7AY, UK\\
\authormark{2}Diamond Light Source Ltd., Diamond House, Harwell Science and Innovation Campus, Didcot, OX11 0DE, UK}

\email{\authormark{*}ot280@bath.ac.uk} %% email address is required



\begin{abstract}
Combinations of spectroscopic analysis and microscopic techniques are used across many disciplines of scientific research, including material science, chemistry and biology. X-ray spectromicroscopy, in particular, is a powerful tool used for studying chemical state distributions at the micro and nano scales. With the beam fixed, a specimen is typically rastered through the probe with continuous motion and a range of multimodal data is collected at fixed time intervals.

The application of this technique is limited in some areas due to: long scanning times to collect the data, either because of the area/volume under study or the compositional properties of the specimen; and material degradation due to the dose absorbed during the measurement. In this work, we propose a novel approach for reducing the dose and scanning times by undersampling the raster data. This is achieved by skipping rows within scans and reconstructing the x-ray spectromicroscopic measurements using low-rank matrix completion. The new method is robust and allows for 5 to 6-fold reduction in sampling. Experimental results obtained on real data are illustrated.
\end{abstract}
%X-ray spectromicroscopy is a powerful tool used for studying material distribution at the microscopic level. Combining spectroscopic analysis with microscopic techniques, it is frequently used across many disciplines of scientific research including material science, chemistry and biology. Despite its benefits, the traditional data collection setting has some significant weaknesses: long scanning times because of the large amount of data that must be collected, and material degradation due to high-energy x-ray radiation. In this work, we propose a novel approach for reducing scanning times by undersampling, reconstructing, and analysing x-ray spectromicroscopic measurements using low-rank matrix completion. The new method allows the selection of robust undersampling ratios and other matrix completion parameters, while minimising the impact of undersampling on the cluster analysis. Results obtained on real data are illustrated.

%%%%%%%%%%%%%%%%%%%%%%%%%%  body  %%%%%%%%%%%%%%%%%%%%%%%%%%
\section{Introduction}

X-ray spectromicroscopy combines x-ray spectroscopy and x-ray microscopy to map changes in chemical state across a specimen on the micro and nano scales. These techniques have been broadly applied to problems across life and physical sciences, such as chemical engineering \cite{application_diesel}, material science \cite{application_batteries}, and biology \cite{application_biology}.
A spectromicroscopy experiment involves measuring an $n_1 \times n_2$ grid at $n_E$ energy levels across the absorption edge of an element of interest.  The resulting $n_1 \times n_2$ spectra represents a big data challenge; to extract meaningful information they are typically analyzed using PCA and cluster analysis to reduce to a mapping of representative spectra, or to a low-rank representation of the data. 

While the technique has been generally successful, the application of spectromicroscopy to in-situ studies and in areas of soft matter or biological materials is limited by two main factors. First, the total experiment time required to collect the data to a given statistical significance; second, the total radiation dose over the collection and any resulting damage to the object, or changes to the chemical state, that may occur as a result. 
The issue of damage due to dose and long collection times occurs across both x-ray and electron optical systems.

To alleviate this issue, a variety of approaches to reduce the number of samples has been proposed. In electron tomography, \emph{compressed sensing} schemes have been investigated to solve the missing wedge problem, or to reduce the number of angles used \cite{Leary2013,Guay2016,Cossa2021}. Random sampling or jittered row sampling has also been used with in-painting to reduce dose and experiment time \cite{Kovarik2016,Browning2018}.  In the x-ray regime, sparse studies are limited, but a low-rank matrix decomposition approach using PCA analysis of spectrotomography datasets has also recently been demonstrated, which merges the angular and energy measurements to reduce overall measurement time \cite{Gao2021}. 

\emph{Low-rank matrix completion} is a well-known inverse problem that was widely used in sensor networks\cite{oh-sensors-2010}, computer vision\cite{ji-video-completion-2010} and medical imaging\cite{liang-spatiotemporal-2007,sherry-MRI-2020}. See e.g. \cite{hansen-inv-problems-2010} for a review of those and further inverse problems. The classical formulation of matrix completion\cite{candes-completion-2009} aims to recover the missing elements of a low-rank matrix from an incomplete set of known entries, each of which is sampled \emph{independently} at random (for example, uniformly).

However, there are important differences between x-ray and electron systems that make independent sampling unattractive for the reduction of the acquisition time in x-ray spectromicroscopy.
An electron beam is rapidly moved across the specimen and can be blanked electrostatically at high rate to control dose, and recreate random patterns. 
In contrast, an x-ray beam is fixed, and the specimen is moved mechanically instead. 
Moreover, a mechanical chopping of the x-ray beam is needed to recreate sampling patterns. 
The mechanical operations in the x-ray regime limit the translation of some of these schemes from equivalent electron experiments.

The use of non-uniform sampling patterns in matrix completion has been widely studied, with the main goal of reducing the approximation error. Non-uniform sampling methods include adaptive cross interpolation\cite{tee-cross-2000}, maximizing the volume of sampled submatrices \cite{gt-maxvol-2001}, adaptive importance sampling using leverage scores\cite{Mahoney-stat-leverage-2012} or application-oriented distributions\cite{Labouesse-adapt-microscopy-2020}, and supervised learning using a training dataset\cite{sherry-MRI-2020}.
However, all of these methods still assume some arbitrary control over the sampling pattern rather than conventional acquisition patterns such as a raster scan.

In this paper we propose a novel approach for undersampling and reconstructing low-rank x-ray spectromicroscopy data that 
uses the \emph{raster sampling pattern}. Our motivation with this work is to deliver a solution that can be deployed routinely at spectromicroscopy facilities by non-experts. This requires an approach where no intervention or tuning is needed to produce results. The method should also work with the standard raster acquisition approaches at these facilities, in particular our initial experimental focus was on optimizing fast low-dose in-situ experiments to study the evolution of battery materials over time. Scanning along a line can be carried out with a faster mechanical movement, hence the time per pixel can be reduced compared to independent or adaptive sampling. In addition, we can lower the x-ray dose on the specimen without compromising the recovery of missing entries: we develop a procedure to generate a robust raster pattern such that each row and column of the matrix has at least one sampled element.

Besides the sampling pattern, the performance of matrix completion depends on the cost function.
In addition to the squared misfit of the sampled elements, the cost function may include regularization terms, promoting sparsity\cite{sherry-MRI-2020} or smoothness\cite{kdhb-neuro-2019}.
Alternatively, the entire completion problem can be turned into a Bayesian inference problem by introducing a prior distribution on the low-rank matrix factors, treated as random matrices, and a likelihood of the observed data samples\cite{Babacan-bayes-lr-2012,Batselier-bayes-tt-2022}.
As a by-product, sparsity-promoting priors may provide an automatic selection of the rank as the number of nonzero posterior components\cite{Hawkins-bayes-tt-2022}.
However, mathematical study of spectromicroscopy is still in some infancy, and lacks well-recognised priors.
The only regularization assumption that is generally valid is the low-rankness of the true absorption distribution.
Therefore, we start with a simple Alternating Steepest Descent algorithm (ASD) \cite{ASD} with the squared misfit cost only.
For a reliable selection of the rank and number of samples, we propose an algorithm, LoopedASD, which successively increases the rank from $1$ to some generous value (e.g. $20$), taking a lower-rank result as the initial guess in each step. This provides a smoother convergence which allows the KNEEDLE \cite{kneedle} algorithm (already used in spectromicroscopy for the PCA analysis) to determine the final rank accurately.

Five datasets were used in this study. The first three (labelled \textit{DS1, DS2, DS3}) are full datasets that can be undersampled numerically after the experiment; the final two (labelled \textit{DS4, DS5}) are measured using both full and sparse raster patterns to verify the experimental implementation. The sparse experimental measurements were conducted at a range of undersampling ratios (details on the different sampling approaches can be found in Section \ref{sec:sampling}). The specimen were produced by mixing $\text{Fe}_2\text{O}_3,\ \text{Fe}_3\text{O}_4$ and FeO powders. The powders were ground in a ball mill and then drop cast onto a silicon nitride membrane.
The dimensions of each data set can be seen in Table \ref{tab:meta_data}. The pairs (DS1, DS2) and (DS4, DS5) are data derived from the same specimen but at different spatial dimensions, i.e. DS2 is DS1 focused on a smaller area, as is DS4 of DS5. We aim to illustrate results using all 5 datasets, but where space is restrictive we only show the 3 independent datasets: DS1, DS3, DS5.

\begin{table}
    \centering
    \caption{Spectromicroscopy Meta Data}
    \label{tab:meta_data}
    \begin{tabular}{c|c c c c}\hline
        Dataset & Data Acquisition & $n_E$ & $n_1$ & $n_2$ \\
        \hline
        DS1 & Full & 149 & 101 & 101\\
        DS2 & Full & 150 & 92 & 79 \\
        DS3 & Full & 152 & 55 & 54\\
        DS4 & Full \& Sparse & 152 & 40 & 40\\
        DS5 & Full \& Sparse & 152 & 80 & 80\\
        \hline
    \end{tabular}
\end{table}

The paper is organised as follows. We begin in Section \ref{sec:low rank model} with a description of the x-ray spectromicroscopy model, and derive its low rank nature. Next, we discuss general matrix completion and raster-aware sampling, which leads into our proposed matrix completion algorithm in Sections \ref{sec:sampling} \& \ref{Sec:completion alg}. Finally, in Sections \ref{sec:Results} \& \ref{sec:sparse scans}, we discuss the results comparing  the reconstructed sparse data against full data.

\section{Low rank model}\label{sec:low rank model}

When the energy of a photon increases beyond the binding energy of a core electron, we see a sharp rise in a material's absorption - an \emph{absorption edge}.  The absorption coefficient, $\mu$,  will vary or be modulated by the local chemical environment. Measuring around the edge, x-ray near edge absorption spectroscopy (XANES) can be used as a fingerprinting tool to identify known standards or materials; investigation of the energy past the absorption edge, the Extended X-ray Absorption Fine Structure (EXAFS), can be used to extract information of nearest neighbour bonding and coordination. Further details regarding the different experimental setup and derivation of the following formula can be found in \cite{XAFS}.

The variation of $\mu$ with energy depends on how an element is chemically bonded and when there are more than one chemical states present the absorption coefficients sum linearly. For a mixture of $S$ materials, the measured x-ray absorption (or optical density), $D(E)$, can be modeled as, 
\begin{equation}\label{eq1:mixed}
    D(E)  =  \sum^S_{s = 1} \mu_s(E)\ t_s,
\end{equation}
where $\mu_s(E)$ and $t_s,\ s = 1,...,S$ are the distinct absorption coefficients ($cm^{-1}$), and thicknesses ($cm$) of the $S$ materials, respectively.

With a focused x-ray beam, this experiment can be performed over $n_1 \times n_2$ positions, and at $n_E$ distinct energies. By stacking each spatial scans, the distribution of x-ray absorption spectra can now be represented as a 3D tensor $D \in \mathbb{R}^{n_E \times n_1 \times n_2}$,
\begin{equation}
    D_{ij_1j_2} = \sum^S_{s = 1} \mu_s (E_i) (t_s)_{j_1j_2}\; ,
\end{equation}
where $(t_s)_{j_1j_2}$ is the thickness of material $s$ at pixel $(j_1,j_2)$, and $\mu_s(E_i)$ is the absorption coefficient of material $s$ at the $i-th$ energy level. This dataset can be flattened by considering a matrix $A \in \mathbb{R}^{n_E \times N}$, such that
\begin{equation}\label{eq: flat model}
    A_{ij} = \sum^S_{s = 1} \mu_s(E_i) t_{sj} + \eta_{ij}\;,
\end{equation}
where $j=1,...,N,\ N = n_1n_2$ indexes over all pixels, and $\eta \in \mathcal{N}(0,\delta^2I)$ represents Gaussian noise with standard deviation $\delta \in \mathbb{R}$.  With a slight abuse of notation, consider the matrices $\mu\in \mathbb{R}^{n_E \times S},\ \mu_{is} = \mu_s(E_i)$ and $t\in \mathbb{R}^{S \times N}$; the columns of $\mu$ represent the absorption coefficients of each material within the specimen, and the rows of $t$ represent the thickness/presence of each material within each pixel. This allows us to write Eq. \eqref{eq: flat model} as 
\begin{equation}
    A = \mu t + \eta.
\end{equation}
This is significant, as it illustrates that for the majority of experiments, where the specimen is made up of only a few materials or chemical states ($S$ is low), the corresponding spectromicroscopy data is inherently \emph{low rank}, as $\text{rank}(\mu t) = S$. With the addition of noise, $A$ is approximately low rank.

To analyse the spatial distribution of the absorption spectra, standard techniques are used to filter and decompose $A$ back to a smaller set of representative spectra; see, for instance, \cite{Cluster_Analysis}.  First, PCA is applied to reduce the noise in the data by producing a rank-$L$ approximation of the most significant components. We compute
\begin{equation}\label{eq:rank L approx}
    A'  = C'R',
\end{equation}
with $A' \in \mathbb{R}^{n_E \times N},\ C' \in \mathbb{R}^{n_E \times L},\ R' \in \mathbb{R}^{L \times N}$. The variable $L$ is chosen to capture as much variation in the data as possible with the smallest rank, and should approximate $S$: it is typically set to be the elbow point (point of maximum curvature) of the singular values of $A$, and can be selected automatically using algorithms like KNEEDLE \cite{kneedle}.

The decomposition in Eq. \eqref{eq:rank L approx} is similar to the model in Eq. \eqref{eq: flat model}, however the columns of $C'$ are abstract spectra - linear combinations of the true spectra with no physical interpretation. Similarly, the rows of $R'$ are the corresponding abstract thickness maps.
Pixels are now clustered together based on the similarity of the normalised mixing factors of the PCA components. This is achieved by clustering the columns of $R'$ using standard clustering algorithms such as kmeans \cite{kmeans} and lvq \cite{LVQ}. Taking the mean spectrum from the columns of $A$ for each cluster increases the signal to noise ratio when compared to the measurements from an individual pixel, and produces accurate x-ray absorption spectra for the dominant material in each cluster.

In \cite{Cluster_Analysis}, it is noted that the first principal component from the PCA (the component with the greatest variation in the data) often describes the average x-ray absorption data across the whole specimen. In some cases the first principal component is discarded, or scaled down, and cluster analysis is only applied to the remaining components. This is done to emphasise the more subtle features and variations in the data and ensure the clustering results are determined by differences in materials not the thickness of the specimen. In this paper, we will refer to the process of discarding the first principal component as Reduction of Thickness Effect (RTE), which is achieved by simply removing the first row from $R'$ in Eq.~\eqref{eq:rank L approx} before applying kmeans to its columns. Generally, the comparison between reconstructed and full data are worse when RTE is used, and they have been applied to several tests to provide the worst-case results. Any cluster results that have used RTE will be noted clearly. Further details on PCA, cluster analysis and RTE can be found in the supplemental document. 

Figure~\ref{fig:schematics} shows a schematic of the sparse spectromicroscopy process. In our proposed scheme, we measure only a small proportion of the data, and recover the missing entries later using low rank matrix completion. The remaining steps are consistent with a standard x-ray spectromicroscopy scheme. Precise details on the sampling are discussed in Section \ref{sec:sampling}, and the completion methods are described in Section \ref{Sec:completion alg}. The image illustrates the data acquisition, the formatting and flattening of the data tensor, the reconstruction of the sparse entries, and the PCA and cluster analysis.

\begin{figure}
    \centering
    \includegraphics[width = \textwidth]{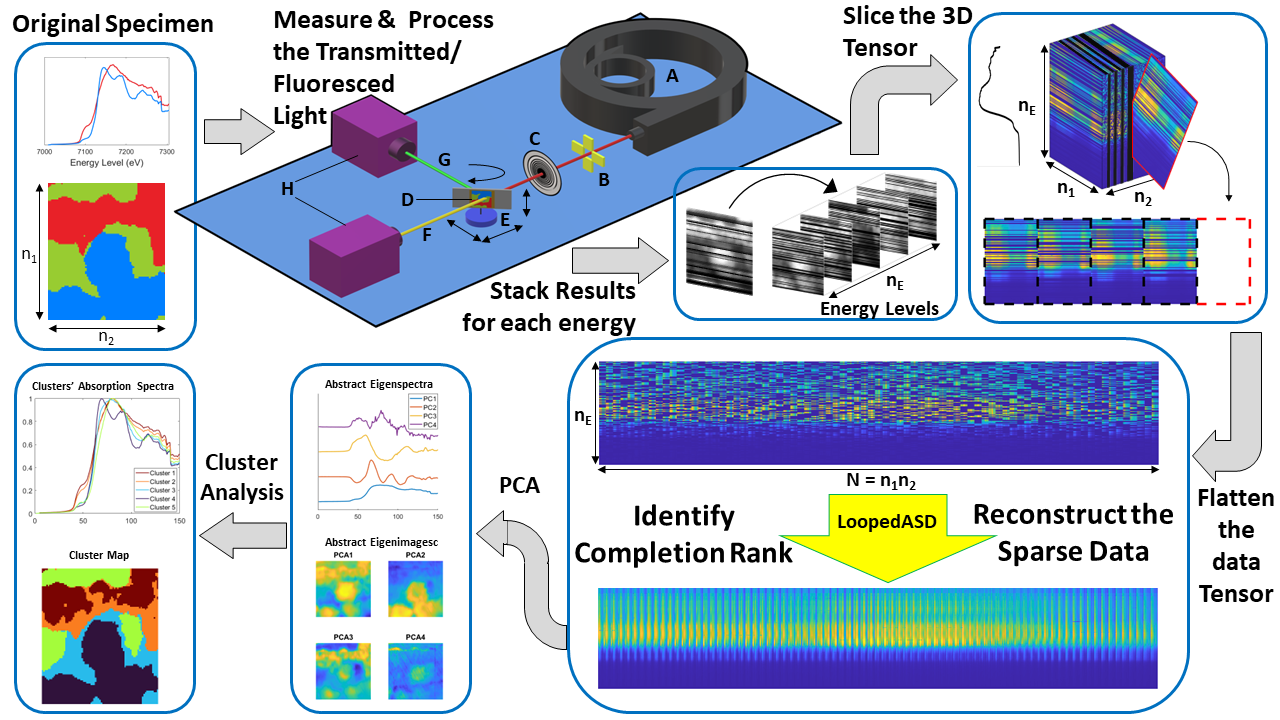}
    \caption{Schematics of Sparse Spectromicroscopy using DS2. The original specimen contains a mixture of two materials: FeO in the blue region and $Fe_2O_3$ in the red region over the background (green region). The experimental setup is as follows: (A) a third/fourth generation synchrotron light source produces a beamline of energy $E_i$. (B) the intensity of the incident beam is measured. (C) the light is focused using an optic such as a zone plate or Kirkpatrick-Baez mirror to produce a micro or nanoscale beam. (D) The specimen is mounted on the stage, (E) which is able to move the specimen across in the XY plane to measure at different positions; the specimen can also be moved in the Z direction to bring the sample to the focus of the beam. Depending on the material and it's thickness, we measure \emph{either} the transmitted flux, (F), \emph{or} the fluoresced flux, (G), by counting photons using different detectors (H). The specimen is moved in a raster pattern; for sparse experiments we only sample $pn_1$ rows for each energy. The processed results for each energy are stacked as seen in the diagram. The tensor is flattened by concatenating slices of the sparse data. The missing entries are recovered using LoopedASD - our low rank matrix completion algorithm. PCA is applied to the completed data to produce the most significant components in the data - the eigenspectra and (reshaped) eigenimages seen in the schematic. The results of the cluster analysis with 5 clusters show similar results to the original specimen, with the locations of the two materials identified, and accurate absorption coefficients for each.}
    \label{fig:schematics}
\end{figure}

\section{Raster-aware sampling patterns}\label{sec:sampling}

To formulate the sampling and reconstruction of x-ray spectromicroscopy data, we set out the following notation. Let $\Omega \subset \{1,...,n_E\} \times \{1,...,N\}$ be the set of known measurements, called the \textbf{sampling pattern}. For a matrix $X \in \mathbb{R}^{n_E \times N}$, we define the \textbf{sampling operator} $\mathcal{P}_\Omega$ as 
\begin{equation}
\mathcal{P}_{\Omega}(X) =
    \begin{cases}
     X_{i,j},           & \text{if } (i,j) \in \Omega,\\
    0,                  & \text{if } (i,j) \notin \Omega.\\
    \end{cases}
\end{equation}
Alternatively, the sampling pattern can be thought of as a binary matrix $\Omega \in \{0,1\}^{n_E \times N}$ with $1$s in the locations of the known entries, and $0$s everywhere else. It is then easy to compute $\mathcal{P}_{\Omega}(X) = \Omega \circ X$, where $\circ$ is the Hadamard product. The key parameter for matrix completion is the \textbf{undersampling ratio}, $p$, the proportion of known entries:
\begin{equation}
    p = \frac{|\Omega|}{n_E N}.
\end{equation}

The standard matrix completion problem involves computing a matrix of minimal rank such that the known entries, indexed by $\Omega$, are equal to the given values. However, this problem is notoriously difficult, and many algorithms will instead seek to solve easier, but provably related, problems; see \cite{convex_matrix_completion,svt,RPCP,SVP}.

For this application, the presence of noise means the spectromicroscopy datasets are only approximately low rank - i.e. there is no exact low rank matrix that would perfectly match the subset of known entries $\mathcal{P}_{\Omega}(A)$. A more useful and efficient approach to reconstruction is to fix the rank, $r$, then solve an optimisation problem to find the best rank-$r$ approximation to the known entries. Thus, to reconstruct the sparse set of measurements, $\mathcal{P}_{\Omega}(A)$, we seek to solve,
\begin{equation}\label{eq:completion problem}
    \min_{Z \in \mathbb{R}^{n_E \times N}} \frac{1}{2}||\mathcal{P}_{\Omega}(A) - \mathcal{P}_{\Omega}(Z)||^2_F \qquad \text{subject to} \qquad \text{rank}(Z) = r.
\end{equation}
Here, we use the frobenius norm $||\cdot ||_F$, defined for $X \in \mathbb{R}^{n_E \times N}$,
\begin{equation}
    ||X||^2_F = \sum_{i,j} X^2_{ij}.
\end{equation}
This approach is generally more effective (and far more computationally efficient) than attempting to solve the standard problem; we must, however, correctly input the completion rank $r$, since completion algorithms work best when $r$ is close to the approximate rank of the full dataset. More details on accurately setting $r$ can be found in the supplementary material.

\subsection{Setting the sampling pattern}

In many applications of low rank matrix completion, it is impossible to predict which entries will be known. To model this, the known entries of the sampling pattern are typically set at random. Usually, \emph{Bernoulli sampling} is used, where each entry is sampled i.i.d. (independent and identically distributed) with probability $p$. 

Unlike other settings, in x-ray spectromicroscopy we have complete control over the data acquisition and can set the scanner to implement specific patterns. However, when formulating the sampling selection, we must consider the physical restrictions of the experiment. For x-ray spectromicroscopy, the specimen is on an XY stage and moves at constant velocity through the beam in a \emph{raster pattern} - scanning across each row in turn before switching to the next energy level and repeating the spatial scan. To maximise the efficiency of the modelled sparse scans, we must ensure the known entries of the sampling pattern are collected together into spatial rows to be scanned. The scanner can then move between known entries, quickly passing over the empty rows.

\begin{figure}
 \centering
    \begin{subfigure}{0.45\textwidth}
        \centering
        \includegraphics[width=0.9\textwidth]{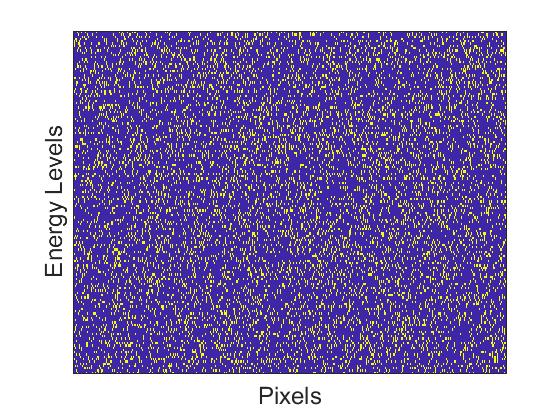}
        \caption{Bernoulli Sampling}
        \label{fig:bernoulli}
    \end{subfigure}
    \begin{subfigure}{0.45\textwidth}
        \centering
        \includegraphics[width=0.9\textwidth]{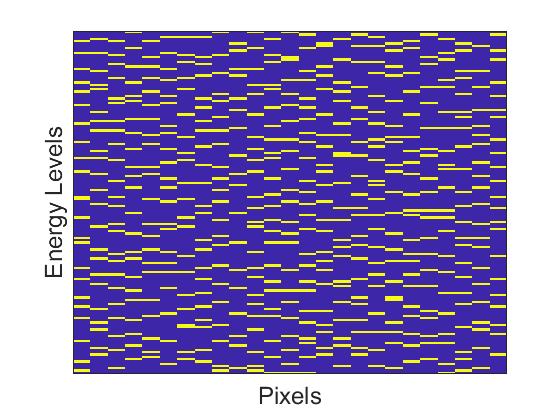}
        \caption{Raster Sampling}
        \label{fig:Raster}
    \end{subfigure}
    \caption{Illustrations of Bernoulli and Raster sampling patterns. Patterns are shown on flattened datasets with $n_E = 150,\ n_1 = 25,\ n_2 = 25,\ p = 0.2$. Yellow indicates the point has been sampled, while purple indicates the entry is missing. }
    \label{fig:Sampling patterns}
\end{figure}

To model the raster aligned patterns, we introduce \emph{Raster sampling}, where spatial rows of the specimen are sampled i.i.d with probability $p$. Illustrations of both Bernoulli and Raster sampling patterns on flattened data sets can be found in Figure \ref{fig:Sampling patterns}, in which the known entries are highlighted yellow. Notice that the raster sampling pattern groups the known entries into the spatial rows of the specimen, which appear as short segments after the data is flattened.

Data that has been sampled with a Bernoulli pattern generally allows completion at lower undersampling ratios, because the known entries are spread more evenly. In practice, undersampling using Bernoulli sampling is possible, however it is only preferable under certain circumstances. Sampling patterns like these can be implemented for x-ray spectromicroscopy in two ways: either the specimen is moved through every position while rapidly blanking the beam, or the specimen is only positioned at the location of each known entry using stop-start motion. The former is difficult to implement and doesn't reduce the experiment time, however may be useful for dose reduction. Conversely, despite the potential use of lower undersampling ratios, stop-start motion is generally less efficient than the continuous motion used for raster sampling, resulting in longer experiments. Indeed, for typical dwell times of around $0.01s$, the time spent accelerating, decelerating and stabilising the scanner outweighs the time saved by scanning fewer entries. Despite this, Bernoulli sampling may be applicable for specimen that require longer dwell times (around $1s$), since reducing the number of known entries has a greater impact on the experiment. Developing methods for such specimen is left for further research.

One issue with Raster sampling is that the grouping of known entries increases the probability that certain rows are not scanned at all, especially for low undersampling ratios. Clearly, it is impossible to recover a row or column of the data set $A$ (distinct from the spatial rows and columns of the specimen) that contains no known entries, so we must ensure that every spatial-row is measured at least once. By testing different sampling patterns, we have also seen that low rank completion is more reliable when the known entries are more evenly spread across the data. 

To promote the spread of data, and ensure every row and column of $A$ contains known entries, we have developed \emph{Robust Raster sampling}. This is a variation of Raster sampling that ensures every row of the specimen is sampled exactly once before any row can be sampled a second time. Further details on Robust Raster Sampling can be found in the supplementary document.

\section{Completion algorithm}\label{Sec:completion alg}

To reconstruct undersampled data, we have developed a low rank matrix completion algorithm \textbf{LoopedASD}. This is a rank-incremental algorithm based on Alternating Steepest Descent (ASD) \cite{ASD}; it uses the output of one ASD-completion as the input for the next higher-rank completion. After testing several different options, \cite{svt,SVP,RPCP}, it was found that this algorithm provided more accurate and reliable results on both simulated data and raster-sampled spectromicroscopy data.

ASD is an iterative method that fixes the rank, $r$, of its iterates by imposing the following decomposition:
\begin{equation}
    Z = XY, \qquad \text{for} \quad Z \in \mathbb{R}^{n_E \times N},\ X \in \mathbb{R}^{n_E \times r},\ Y\in \mathbb{R}^{r \times N}.
\end{equation}
Thus, the optimisation of the problem in Eq.~\eqref{eq:completion problem} now becomes,
\begin{equation}\label{eq:ASD Problem}
    \min_{X,Y}f(X,Y),\quad \text{where} \quad f(X,Y) = \frac{1}{2}||\mathcal{P}_{\Omega}(A) - \mathcal{P}_{\Omega}(XY)||^2_F,\quad X \in \mathbb{R}^{n_E \times r},\quad Y \in \mathbb{R}^{r \times N}.
\end{equation}
We minimise the function $f$ by alternately optimising the components $X$ and $Y$ using steepest descent with exact step sizes. Indeed, by fixing one component the gradient of $f$ can be computed easily. It is then possible to analytically compute the exact step size required to minimise $f$ along the gradient direction. Once the factor has been updated, we alternate the fixed component and repeat the process, iterating until suitable stopping conditions are satisfied.

Let $f(X,Y)$ be written as $f_Y(X)$ for fixed $Y$ and $f_X(Y)$ for fixed $X$; the corresponding gradients are written $\nabla f_Y(X)$ and $\nabla f_X(Y)$, and the exact step sizes are written $\eta_{X},\ \eta_Y$. Beginning with random matrices $X_0 \in \mathbb{R}^{n_E \times r},\ Y_0 \in \mathbb{R}^{ r \times N}$, we implement ASD as follows,

\begin{equation}
    \begin{cases}
        \text{Fix}\ Y_i,\ \text{compute}\ \nabla f_{Y_i}(X_i)\ \&\ \eta_{X_{i}}\\
        X_{i+1} = X_{i} - \eta_{X_{i}}\nabla f_{Y_i}(X_i)\\
        \text{Fix}\ X_{i+1},\ \text{compute}\ \nabla f_{X_{i+1}}(Y_i)\ \&\ \eta_{Y_{i}}\\
        Y_{i+1} = Y_{i} - \eta_{Y_{i}}\nabla f_{X_{i+1}}(Y_i).
    \end{cases}
\end{equation}
The gradients and step sizes are given below. Full derivations can be found in the supplemental documents:

\begin{align}
    \nabla f_Y(X) &= -(\mathcal{P}_{\Omega}(A) - \mathcal{P}_{\Omega}(XY^T))Y, \qquad  \qquad &\nabla f_X(Y) &= -X^T(\mathcal{P}_{\Omega}(A) - \mathcal{P}_{\Omega}(XY^T)).\\
    \eta_X  &= \frac{||\nabla f_Y(X)||^2_F}{||\mathcal{P}_{\Omega}(\nabla f_Y(X)Y^T)||^2_F}, &\eta_Y &= \frac{||\nabla f_X(Y)||^2_F}{||\mathcal{P}_{\Omega}(X[\nabla f_X(Y)]^T)||^2_F}
\end{align}
One advantage of ASD is that the residual $(\mathcal{P}_{\Omega}(A) - \mathcal{P}_{\Omega}(XY))$ can be easily updated between iterations, removing the need to compute the matrix product $XY$ for each iteration. Thus, the per iteration cost of ASD has leading order $8|\Omega|r$ (see \cite{ASD}). 

A maximum number of iterations was set as a stopping condition, as well as a tolerance on the relative norm of the residual at each iteration, 
\begin{equation}\label{eq:resnorm}
\frac{||\mathcal{P}_\Omega(A) - \mathcal{P}_\Omega(X_i Y_i)||_F^2}{||\mathcal{P}_\Omega(A)||^2_F}. 
\end{equation}
Once either stopping condition is satisfied, the two factors are output by the algorithm and are denoted $X^*$ and $Y^*$.

To evaluate the success of the completion algorithms, we compute the \emph{completion error}, often denoted $e_c$. This is simply the relative norm of the difference between the true low rank matrix $A$ and the output of the ASD algorithm $A^* = X^*Y^*$, defined as,
\begin{equation}\label{eq:ec}
    e_c = \frac{||A-A^*||^2_F}{||A||^2_F}
\end{equation}
A more detailed description of the algorithm can be found in the supplemental material.
\subsection{Improving results with LoopedASD}

Following rigorous testing of ASD for raster sampling, it was found that there exists an approximate optimal undersample ratio, $p^*$, that depends linearly on, and is correlated with, the data's rank $r$. For $p > p^*$, the probability of a successful completion was almost certain, and the completion errors were consistently low. For $p < p^*$, the completion errors increased and the completion rate (the proportion of tests that successfully recover data to a certain accuracy - usually $10^{-4}$) decreased rapidly to zero. Intuitively, the dependence of $p^*$ on $r$ makes sense: to reconstruct more complex datasets (with higher ranks), more known entries are required to successfully recover the missing ones. 

\emph{LoopedASD} was developed to take advantage of this heuristic: beginning with $r = 1$, we use the outputs of the $r = j$ completion as initial guesses for the $r = j+1$ completion, iterating up to the completion rank set by the user.

Let $(X_0)^{(j)} \in \mathbb{R}^{n_E \times j},\ (Y_0)^{(j)} \in \mathbb{R}^{j \times N}$ be the initial matrices for the $j^{th}$ rank step, and $(X^*)^{(j)} \in \mathbb{R}^{n_E \times j},\ (Y^*)^{(j)} \in \mathbb{R}^{j \times N}$ be the outputs of the $j^{th}$ iteration of ASD with completion rank $r=j$. In order ensure the dimensions of the factors increase with each rank increment, we simply concatenate the $j^{th}$ outputs $(X^*)^{(j)},\ (Y^*)^{(j)}$ with a random column and row respectively to produce $(X_0)^{(j+1)},\ (Y_0)^{(j+1)}$.

The aim of this variation is to allow the iterates to converge quickly to a rank-1 approximation of the sparse data, where fewer known entries are required. Once a rank-$j$ approximation is known, the distance to the rank-$(j+1)$ solution should be relatively low, and again fewer known entries are required to converge quickly to the next one. Ensuring iterates remain close to the minimum should also avoid convergence to a spurious local minima of $f$ (which is non-convex).

The completion results of LoopedASD confirm the validity of this heuristic, as it yields good  reconstructions more reliably at low undersampling ratios.

In addition to the usual stopping conditions, an early stopping procedure was required due to slow convergence rates when the iterates approached optimal solutions. In the case that the average change in the norm of the residual (Eq. \eqref{eq:resnorm}) over 50 iterations drops below a second tolerance (typically $10^{-5}$), then the early stopping condition is satisfied and will halt the algorithm.

It has been noted that the completion rank must be set before implementing ASD and LoopedASD. This, however, is not an obvious choice since we cannot evaluate the accuracy of the different rank-$k$ completions without prior knowledge of the results. To overcome this, we have developed a method that implements short completions over several ranks and evaluates the completion errors using cross validation so that we can efficiently identify the optimal completion rank to use. A full description and justification of the method can be found in the supplemental material.

\section{Validation of the algorithm with numerical undersampling}\label{sec:Results}

We test the method in two distinct ways. First, we numerically sample full datasets. To produce these datasets, `unknown' entries are set to zero according to a randomly generated robust raster sampling pattern. Then, in Section \ref{sec:sparse scans}, we implement sparse scans on the beamline to determine if any additional issues arise and to test in a new setting outside of our test set.

To properly evaluate the performance of the methods described in this paper, several measures of success must be considered. We can compare the original and reconstructed datasets directly by computing the average completion error. Perhaps more significantly, we evaluate the difference between the cluster maps and the absorption spectra themselves. Finally, we can plot and visually compare the clusters and absorption spectra. Due to specimen drifts (described in Section \ref{sec:sparse scans}), the first two `computational' approaches can only be achieved using numerically sampled data; visual comparisons must be used when evaluating reconstructions using the sparse scan data. During these tests, we use the optimal completion rank $r^*$ determined by the rank selection algorithm described in the supplemental document to produce the completed matrix $(A^*)^{(r^*)} = (X^*)^{(r^*)} (Y^*)^{(r^*)}$. $A^*$ is then used instead of the full data set $A$ in the analytic processes (PCA and clustering).

\newpage
\subsection{Completion errors of LoopedASD}

\begin{figure}
    \begin{subfigure}{0.33\textwidth}
        \centering
        \includegraphics[width = \textwidth]{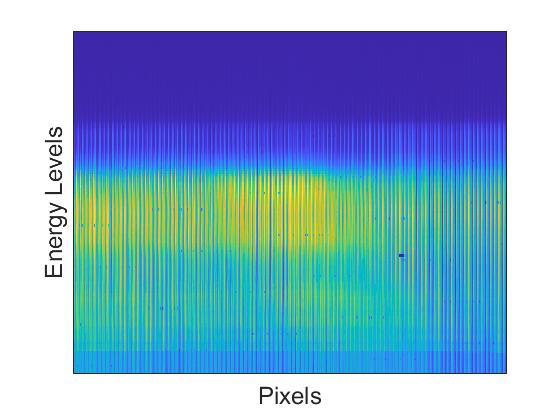}
        \caption{DS1 Original}
        \label{fig:DS1 original}
        \vspace{0.33cm}
    \end{subfigure}
    \begin{subfigure}{0.33\textwidth}
        \centering
        \includegraphics[width = \textwidth]{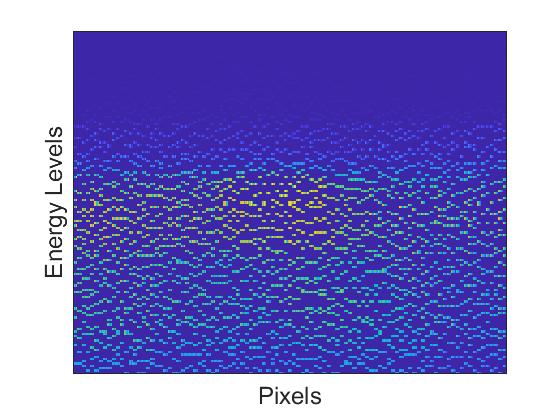}
        \caption{DS1 Sampled}
        \label{fig:DS1 Sampled}
        \vspace{0.33cm}
    \end{subfigure}
        \begin{subfigure}{0.33\textwidth}
        \centering
        \includegraphics[width = \textwidth]{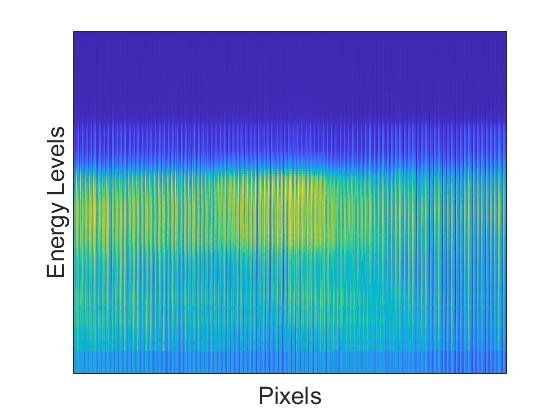}
        \caption{\makecell{DS1 Completed. $r = 6$, \\ $e_c= 0.048,\ \min(e_c) = 0.039$.}}
        \label{fig:DS1 completed}
    \end{subfigure}
    
    \begin{subfigure}{0.33\textwidth}
        \centering
        \includegraphics[width = \textwidth]{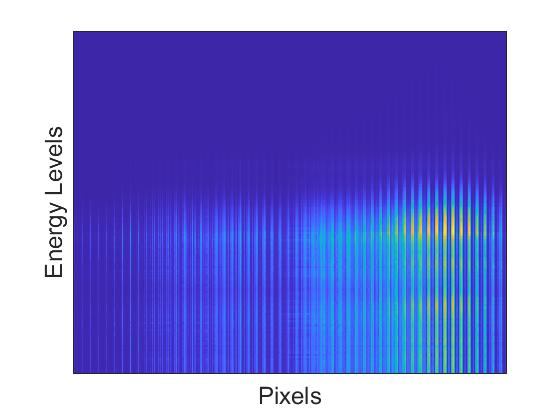}
        \caption{DS3 Original}
        \label{fig:DS3 original}
        \vspace{0.33cm}
    \end{subfigure}
        \begin{subfigure}{0.33\textwidth}
        \centering
        \includegraphics[width = \textwidth]{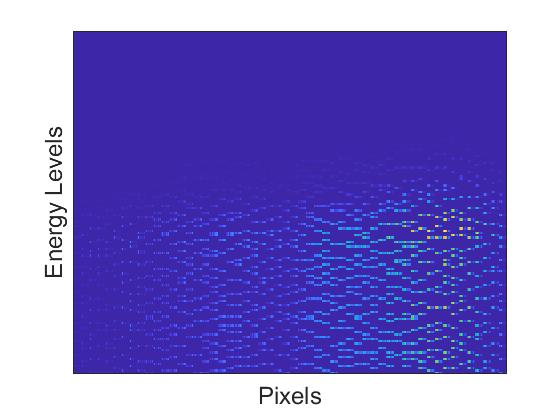}
        \caption{DS3 Sampled}
        \label{fig:DS3 Sampled}
        \vspace{0.33cm}
    \end{subfigure}
    \begin{subfigure}{0.33\textwidth}
        \centering
        \includegraphics[width = \textwidth]{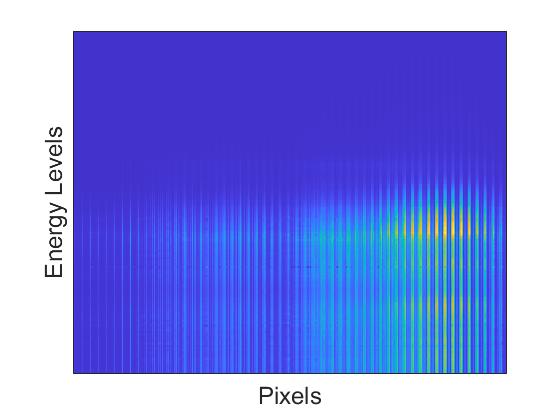}
        \caption{\makecell{DS3 Completed. $r = 4$, \\ $e_c = 0.052,\ \min(e_c) = 0.037$.}}
        \label{fig:DS3 completed}
    \end{subfigure}
    
    \begin{subfigure}{0.33\textwidth}
        \centering
        \includegraphics[width = \textwidth]{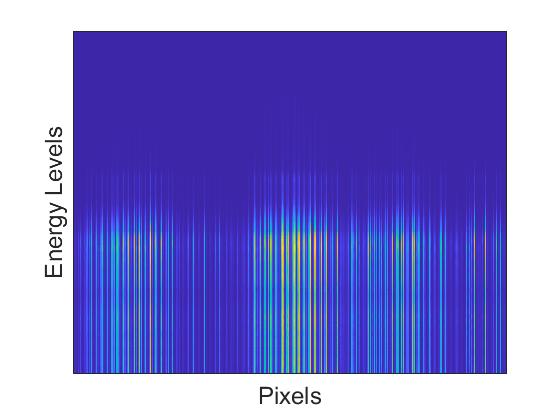}
        \caption{DS5 Original}
        \label{fig:DS5 original}
        \vspace{0.33cm}
    \end{subfigure}
    \begin{subfigure}{0.33\textwidth}
        \centering
        \includegraphics[width = \textwidth]{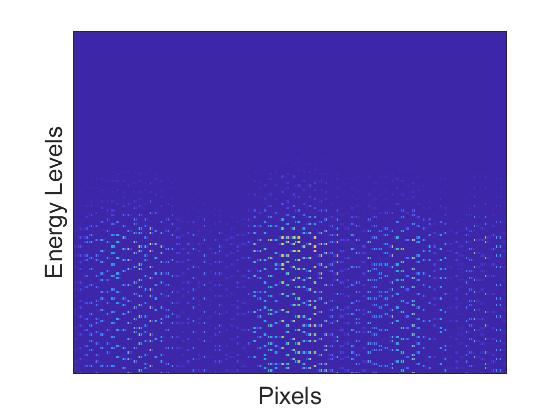}
        \caption{DS5 Sampled}
        \label{fig:DS5 Sampled}
        \vspace{0.33cm}
    \end{subfigure}
    \begin{subfigure}{0.33\textwidth}
        \centering
        \includegraphics[width = \textwidth]{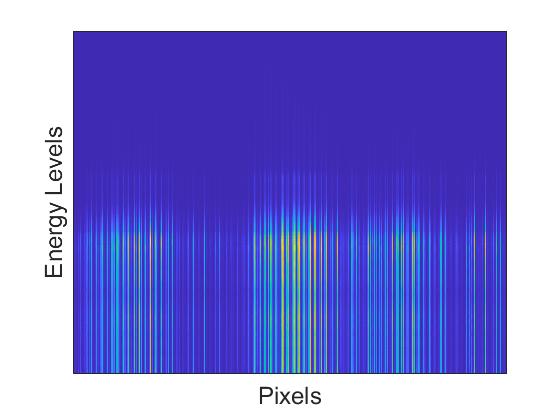}
        \caption{\makecell{DS5 Completed. $r = 5$, \\ $e_c = 0.021,\ \min(e_c) = 0.014$.}}
        \label{fig:DS5 completed}
    \end{subfigure}
    \caption{Reconstructions of the three independent datasets, using LoopedASD as the completion algorithm. Each figure is a scaled colour image of the flattened dataset, with energy levels across the vertical axis, and all pixels across the horizontal axis. The left column shows the original datasets. The middle column shows the sampled data at $p=0.20$. The right column shows the reconstructed data. Below each reconstruction is the completion rank, $r$, the completion error $e_c$ (see Eq. \eqref{eq:ec}), and the minimum approximation error (see Eq. \eqref{eq:EYM Thm}) for that completion rank.}
    \label{fig:LoopedASD illustration}
\end{figure}

We first examine the norm of the difference between data sets. When conducting experiments such as this, it is important to consider the constraints on the completion error, $e_c$ (Eq. \eqref{eq:ec}), for a rank-$r$ completion. 

Consider a rank-$r$ approximation of a matrix $A$, denoted $A^{(r)}$. We can compute the minimum approximation error of $A^{(r)}$ in the Frobenius norm using the Eckart-Young-Mirsky (EYM) Theorem \cite{E-Y-M_Theorem}. We shall refer to this value as the minimal rank-$r$ approximation error, and it is given by:
\begin{equation}\label{eq:EYM Thm}
    \min_{\text{rank}(A^{(r)}) = r} ||A - A^{(r)}||_F = \sqrt{\sigma_{r+1}^2 + \sigma_{r+2}^2 + ... + \sigma_{n}^2},
\end{equation}
where $\sigma_1,\ ...\ ,\ \sigma_n$ are the singular values (SVs) of $A$. Since any completion result with a completion rank of $r$ is just an example of a rank-$r$ approximation, the corresponding minimum approximation errors are lower bounds for the corresponding completion errors. Thus the success of any completion should be evaluated by comparing back to the corresponding minimum error.

LoopedASD produces successful completions for all datasets, and for undersampling ratios as low as $p=0.15$. By examining the mean completion errors, $e_c$, one can identify the optimal undersampling ratio, $p^*$. As before, for $p < p^*$ average completion errors rise quickly, and for $p > p*$ average completion errors decrease slowly, reducing the benefit of taking further measurements. It was found that LoopedASD produces lower completion errors more reliably than ASD, in particular for undersampling ratios around $p^*$.

To help with the visualization of the completion, we plot the flattened data sets themselves as colourmap images. Purple points indicate a value of zero, while bright yellow points indicate higher measured values. In Figure \ref{fig:LoopedASD illustration} we provide a visual representation of the sampling and reconstruction process at the optimal rank, $r^*$, and with a consistent undersampling ratio of $p=0.20$. Note that completions often appear smoother than the original data - since it is low rank, it will have filtered out much of the noise.

\begin{figure}
    \centering
    \begin{subfigure}{0.49\textwidth}
        \centering
        \includegraphics[width = 0.85\textwidth]{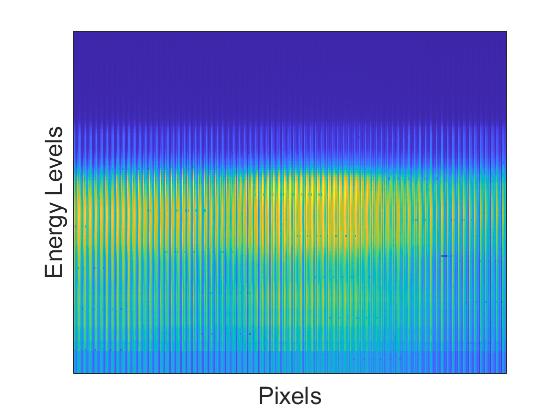}
        \caption{DS2 Original\\  \textcolor{white}{a} }
    \end{subfigure}
        \begin{subfigure}{0.49\textwidth}
        \centering
        \includegraphics[width = 0.85\textwidth]{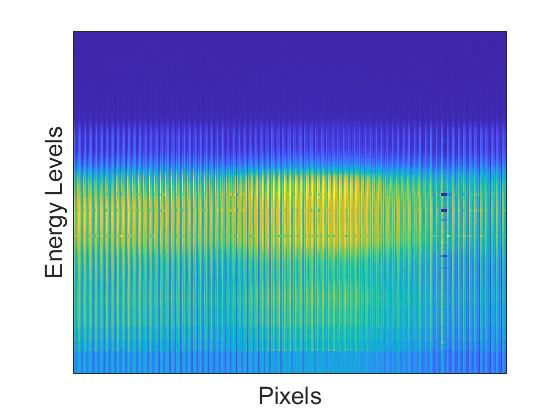}
        \caption{DS2 Completed, $r = 5$,\\ $e_c = 0.52,\ \min(e_c) = 0.026$}
    \end{subfigure}
    \caption{Completion result following intentional sampling of the artifact in DS2 (dark spot towards the right). Note that the corrupted points are contained to the column of the initial artefact.}
    \label{fig:error localisation}
\end{figure}

One advantage of ASD and LoopedASD is that the impact of any artifacts that have been sampled remain contained in their rows/columns in the dataset. In Figure \ref{fig:error localisation} we illustrate this by intentionally sampling a corrupted entry in DS2. We can see that the majority of the image remains accurate and it is just the row and column that contained the original artefact that are affected. Indeed, it is very easy to identify the corrupted regions, which can be cut from the dataset. The example in Figure \ref{fig:error localisation} initially had a completion error of $0.052$, compared to the mean $0.33$ (the corresponding minimum approximation error is $0.026$). Once the affected entries are removed, $e_c$ is computed as $0.047$ providing a much stronger result. In practice this property is very useful, since it implies the process is robust against artefacts found in the data.
\subsection{Impact of completion on cluster analysis}

We now compare the results of the cluster analysis from the full data and the sparse data. In some sense, this comparison is more relevant than simply comparing the completion errors because the success of the new methods will be determined by the quality of the cluster maps produced by the reconstructions, not necessarily the abstract difference between the two datasets. 

We can evaluate the similarity of clusters using the Adjusted Rand Index (ARI) \cite{ARI}. This is a symmetric score, with values between $-1$ and $1$. An ARI score of $1$ indicates clusters are a perfect match, while a score of $0$ implies classifications have effectively been assigned at random. Generally we use the inverted score of $(1-\text{ARI})$ to evaluate cluster quality so that $0$ indicates a perfect match, and smaller scores are generally better.

On the other hand, we can compare individual absorption coefficients using the Euclidean 2-norm, but we must ensure that we are taking the difference between coefficients for equivalent materials/clusters. Suppose we have performed cluster analysis on both the full data set and the completed data with $N_{\text{cluster}}$ many cluster centres. Let $\mu_f \in \mathbb{R}^{n_E \times N_{\text{cluster}}}$ denote the absorption coefficients of the full data and $\mu_c \in \mathbb{R}^{n_E \times N_{\text{cluster}}}$ denote those from the completed data. In both cases, the $i^{th}$ columns, $(\mu_f)_i,\ (\mu_c)_i$ respectively, represent the absorption coefficient corresponding to the $i^{th}$ cluster. Note that the $i^{th}$ clusters from the full data may not correspond to the $i^{th}$ cluster from the completed data. We now align the clusters by taking each full-data cluster and finding the completed cluster that best fits it (minimises $(1-\text{ARI})$); using this permutation index, we rearrange the columns of $\mu_c$ to get the aligned absorption coefficients, $\mu_a \in \mathbb{R}^{n_E \times N_{\text{cluster}}}$. We can now be sure the absorption coefficients represent the same area and the same material. We now wish to combine the relative differences of each cluster such that each cluster is weighted equally (not all absorption spectra are of the same magnitude) and is normalised for the number of clusters used. Thus, we compute the \emph{spectral difference, $d_{\text{spec}}$}, as:
\begin{equation}\label{eq:spec_diff}
    d_{\text{spec}} = \frac{1}{\sqrt{N_{\text{clusters}}}}\sqrt{\sum_{i = 1}^{N_{\text{clusters}}} \left( \frac{||(\mu_f)_i - (\mu_a)_i ||_2}{||(\mu_a)_i||_2 } \right)^2}.
\end{equation}

One complication is that cluster analysis is not deterministic, since there is some randomness in the starting vectors of the cluster centres. Because of this, we often find several common cluster maps resulting from the same dataset. This is true for both the full datasets and the reconstructed datasets, which can produce false ARI scores and spectral differences when opposing outputs are compared. To overcome this, we apply a clustering process to the clusters themselves - grouping similar maps together into `Super Clusters'. Taking the mode within each super cluster produces a series of representative cluster maps that are produced by the original data. We can now compare the reconstructed cluster against each representative in turn, recording the highest score as the most appropriate comparison.

The spectromicroscopy datasets were sampled numerically, their optimal ranks were computed, and the data reconstructed using LoopedASD with early stopping. The completion ranks were set to be the optimal rank, $r^*$, using the method described in the supplementary document. PCA is used to decompose the data into its most significant components (we set the number of components used $L = r^*$, so the rank is consistent). Finally, we perform the cluster analysis using kmeans with 5 clusters and compute the associated absorption spectra. For DS1, DS2, DS3 we used RTE (dropping the first principal component before applying kmeans) to provide the worst-case results. Note that if RTE had not been used, both the ARI scores and the spectral difference would improve. For DS4 and DS5, we avoided using RTE due to the poorer quality of the clustering results for both the full and reconstructed data. When using RTE on DS4 and DS5, the clusters are not smooth, the images and spectra were noisy, and the outcomes were not representative of typical results.

\begin{figure}
    \begin{subfigure}{0.49\textwidth}
        \centering
        \includegraphics[width = 0.9\textwidth]{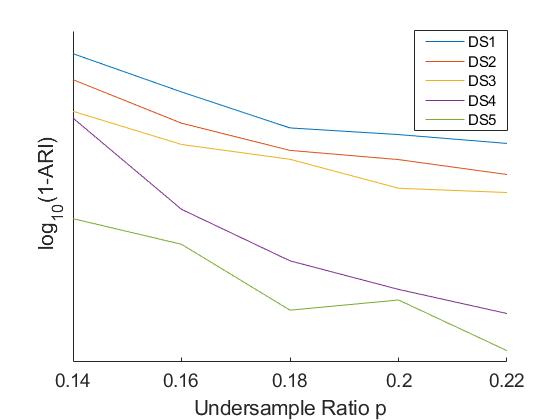}
        \caption{$\log_{10}(1-ARI)$ against undersampling ratio, $p$}
        \label{fig:ARI_vs_p}
    \end{subfigure}
    \begin{subfigure}{0.49\textwidth}
        \centering
        \includegraphics[width = 0.9\textwidth]{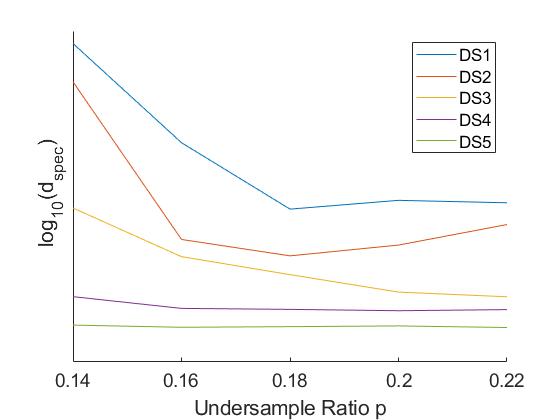}
        \caption{$\log_{10}(d_\text{spec})$ against undersample ratio, $p$}
        \label{fig:SpecDiff_vs_p}
    \end{subfigure}
    \caption{Comparing $\log_{10}$ of the cluster results against undersampling ratio using the optimal completion rank. In each figure, \emph{lower scores are better}. Note that the individual plots have been translated vertically to improve the clarity of the image.}
    \label{fig:Cluster_vs_p}
\end{figure}

Averaging over many iterations, we plot the clustering results for each dataset and for various undersampling ratios in Figure \ref{fig:Cluster_vs_p}. In Figure \ref{fig:ARI_vs_p} we plot $\log_{10}(1-\text{ARI})$ against the undersampling ratio, and in Figure \ref{fig:SpecDiff_vs_p} we plot the log of the spectral difference, $log_{10}(d_\text{spec})$, against the undersampling ratio. The purpose of the plots is to illustrate the qualitative behaviour of the results, so have translated the individual plots vertically to improve clarity. The quantitative results can be found in Table \ref{tab:Sparse Clustering results}. Here we see that, as we increase the undersampling ratio, the $(1-\text{ARI})$ scores and spectral differences behave similarly to the completion error, $e_c$ (Eq. \eqref{eq:ec}). Indeed, by plotting the logs of these scores we see the similarly shaped curves - a faster decline for lower $p$ that flattens for higher $p$. Once again, we identify the optimal undersampling ratio, $p^*$, at the elbow of the plots - the points of maximum curvature. It is clear from the plot where the majority of the optimal undersampling ratios lie, however for DS2 and DS4 the elbows occur at different points across the two plots ($p = 0.16$ for Figure \ref{fig:ARI_vs_p} and $p=0.18$ for Figure \ref{fig:SpecDiff_vs_p}). In these cases, we take the higher value to ensure enough entries are known for better reliability. For DS4 and DS5, the pattern in Figure \ref{fig:SpecDiff_vs_p} is less clear and appears more as a constant value. This is simply because the variation in spectral difference over this range of undersample ratios is much smaller than for the other datasets, and by appropriately scaling these plots one can see the same curve as before.

In Table \ref{tab:Sparse Clustering results}, we record the optimal completion rank, the corresponding optimal undersampling ratio (found at the elbow of the results in Figure \ref{fig:Cluster_vs_p}), the completion errors, $e_c$, and the cluster results produced by these parameters. 

\begin{table}
    \centering
    \caption{Optimal Clustering Results for numerically sampled Spectromicroscopy datasets. For the metrics used ($e_c$,\ $1-\text{ARI}$,\ $d_\text{spec}$), \emph{lower scores are better}.}
    \label{tab:Sparse Clustering results}
    \begin{tabular}{c|c|c|c|c|c|c}\hline
         \textbf{dataset} & \makecell{Optimal \\ Completion\\ Rank, $r^*$} & \makecell{Optimal \\ undersampling \\ ratio, $p^*$} & \makecell{Mean \\ Completion \\ Error \eqref{eq:ec}} & \makecell{RTE \\ used} & \makecell{Mean \\ (1-ARI)} & \makecell{Mean Spectral \\ Difference  \\ \eqref{eq:spec_diff}} \\
         \hline
         DS1 & 6 & 0.18 & 0.0483 & Yes & 0.106 & 0.082\\
         \hline
         DS2 & 5 & 0.18 & 0.0326 & Yes & 0.108 & 0.115\\
         \hline 
         DS3 & 4 & 0.20 & 0.0515 & Yes & 0.207 & 0.509\\
         \hline
         DS4 & 7 & 0.18 & 0.025 & No & 0.009 & 0.096\\
         \hline
         DS5 & 5 & 0.18 & 0.027 & No & 0.006 & 0.152 \\
         \hline
    \end{tabular}
\end{table}

\begin{figure}[H]
    \begin{subfigure}{0.33\textwidth}
        \centering
        \includegraphics[width = 0.8\textwidth]{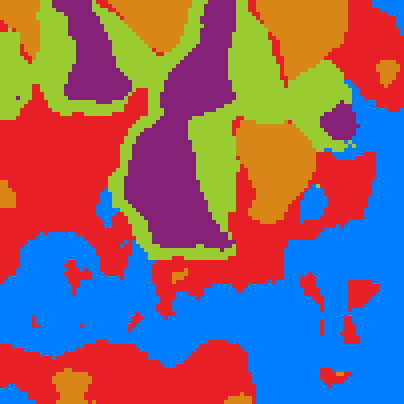}
        \caption{DS1}
        \label{fig:DS1 cluster}
    \end{subfigure}
    \begin{subfigure}{0.33\textwidth}
        \centering
        \includegraphics[width = 0.93\textwidth]{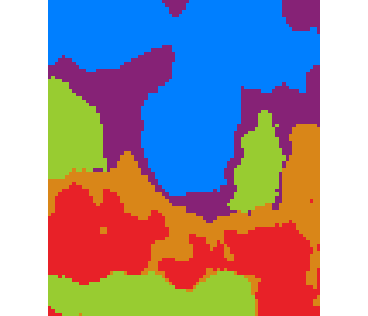}
        \caption{DS2}
        \label{fig:DS2 cluster}
    \end{subfigure}
    \begin{subfigure}{0.33\textwidth}
        \centering
        \includegraphics[width = 0.8\textwidth]{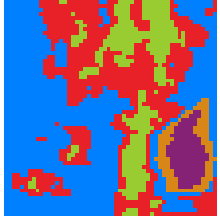}
        \caption{DS3}
        \label{fig:DS3 cluster}
    \end{subfigure}

    \begin{subfigure}{0.33\textwidth}
        \centering
        \includegraphics[width = \textwidth]{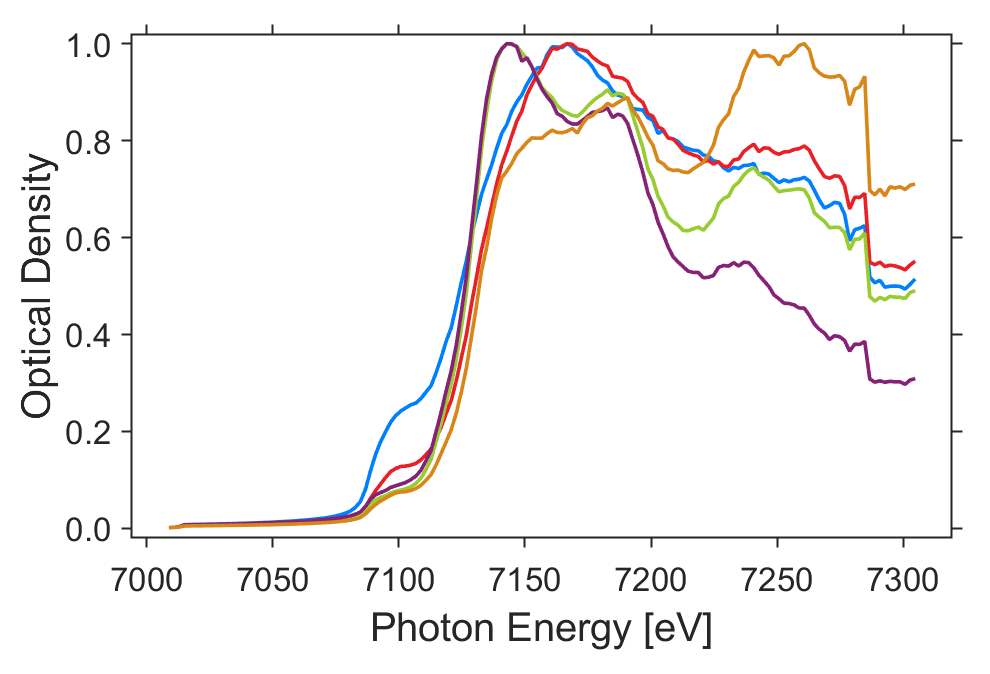}
    \end{subfigure}
    \begin{subfigure}{0.33\textwidth}
        \centering
        \includegraphics[width = \textwidth]{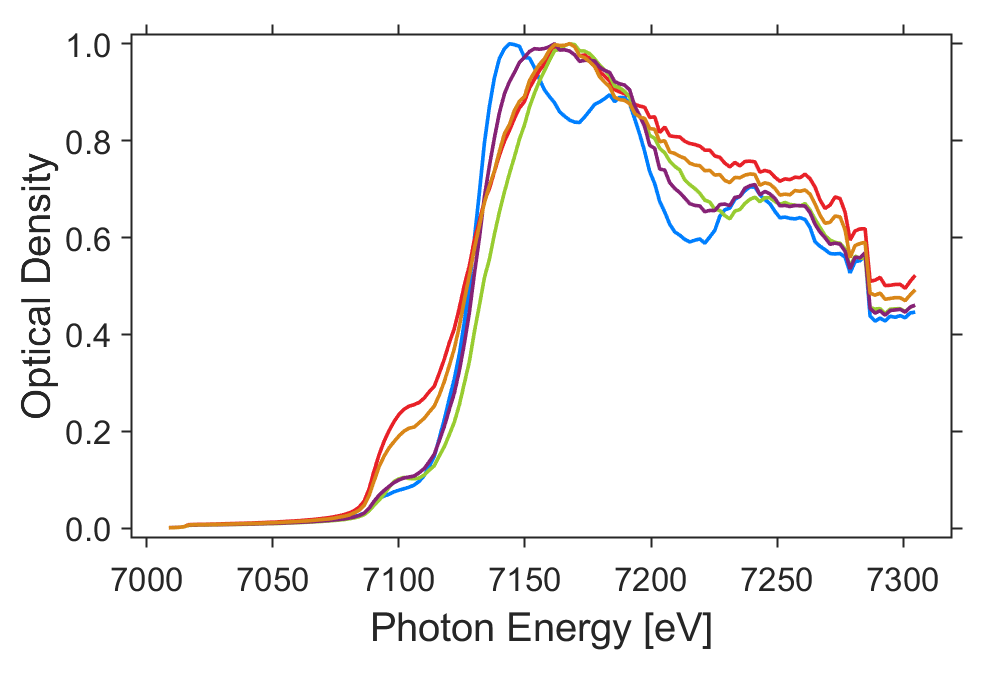}
    \end{subfigure}
    \begin{subfigure}{0.33\textwidth}
        \centering
        \includegraphics[width = \textwidth]{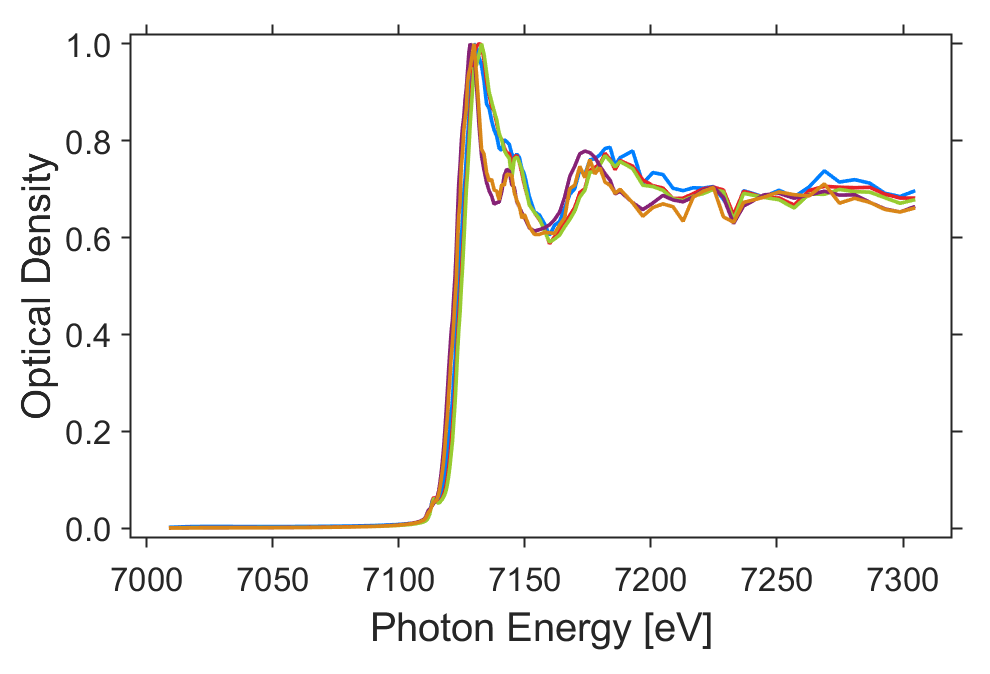}
    \end{subfigure}
    \caption{Cluster and spectral results from full datasets using RTE. Results computed using Mantis-xray \cite{mantis}.}
    \label{fig:full data clusters}
    \vspace{0.02\textheight}

    \begin{subfigure}{0.33\textwidth}
        \centering
        \includegraphics[width = 0.8\textwidth]{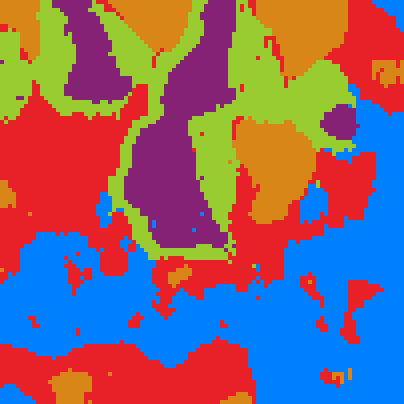}
        \caption{DS1}
        \label{fig:DS1 complete cluster}
    \end{subfigure}
    \begin{subfigure}{0.33\textwidth}
        \centering
        \includegraphics[width = 0.93\textwidth]{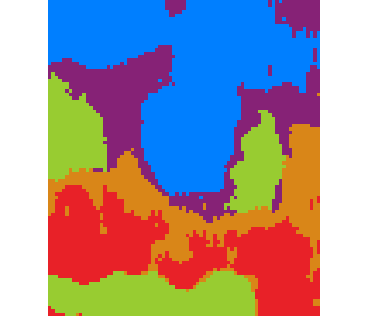}
        \caption{DS2}
        \label{fig:DS2 complete cluster}
    \end{subfigure}
    \begin{subfigure}{0.33\textwidth}
        \centering
        \includegraphics[width = 0.8\textwidth]{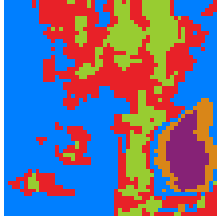}
        \caption{DS3}
        \label{fig:DS3 complete cluster}
    \end{subfigure}

    \begin{subfigure}{0.33\textwidth}
        \centering
        \includegraphics[width = \textwidth]{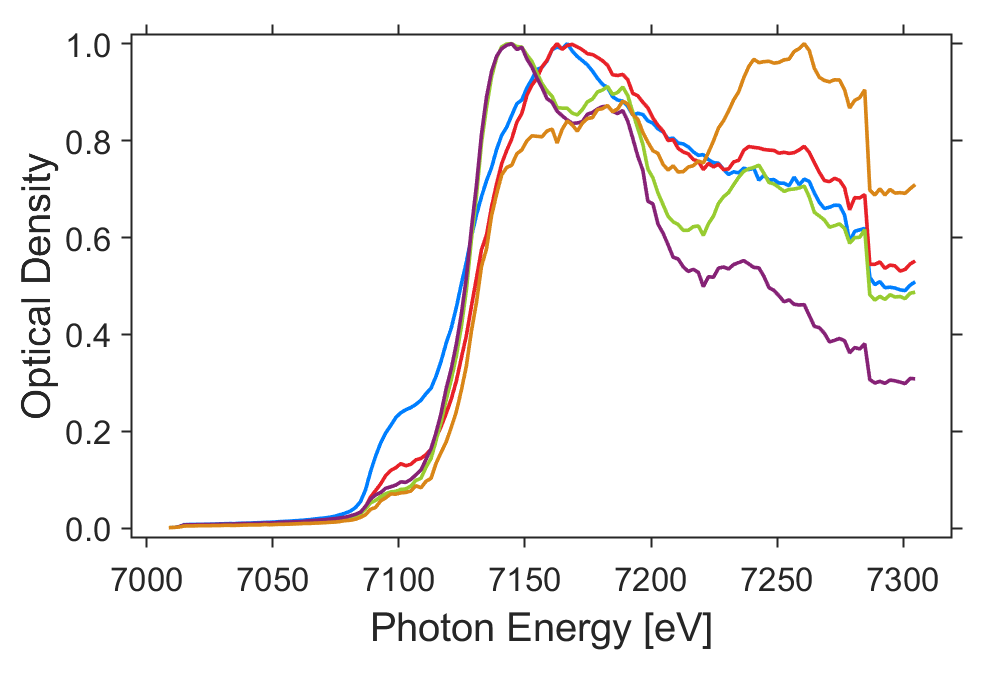}
    \end{subfigure}
    \begin{subfigure}{0.33\textwidth}
        \centering
        \includegraphics[width = \textwidth]{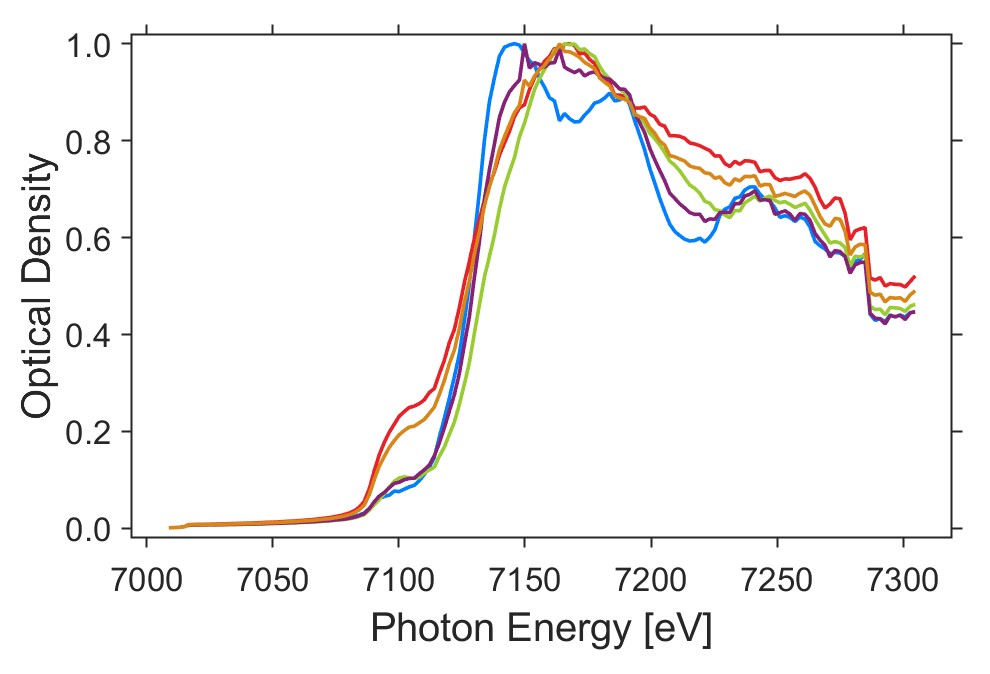}
    \end{subfigure}
    \begin{subfigure}{0.33\textwidth}
        \centering
        \includegraphics[width = \textwidth]{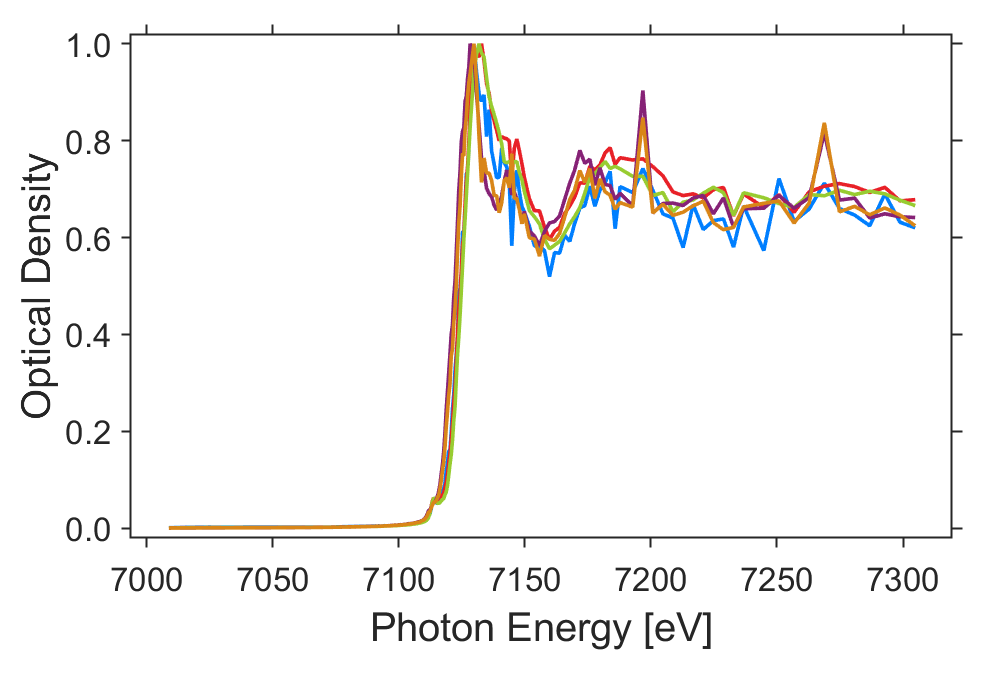}
    \end{subfigure}
    \caption{Cluster and spectral results from reconstructions of sampled data, with $p = 0.20$, using RTE. Results computed using Mantis-xray \cite{mantis}. Please note that for DS1 and DS2 there is a sharp downward going discontinuity at around $7270eV$ in both the full and completed results. This feature is present in the original data, and can be seen in the absorption spectra of many of the pixels with stronger signals. It is likely due to an error in the calibration of the energy stepping, which requires coordinated motion of an insertion device and monochromator, resulting in a strong variation in the intensity of the beam which was not normalised correctly.}
    \label{fig:completion clusters}
\end{figure}

In Figures \ref{fig:full data clusters} \& \ref{fig:completion clusters}, we visually compare the cluster maps and absorption spectra of the full data and the reconstructed data for DS1, DS2 and DS3 (DS4 and DS5 are used in Section \ref{sec:sparse scans} to illustrate the sparse scans). The data sets are sampled numerically with $p = 0.20$, the optimal completion rank was used, and each of these results uses RTE. These images were produced by Mantis X-ray \cite{mantis}, an open source software package used for analysing x-ray spectromicroscopy data.

We can see that both the cluster maps and absorption spectra are very similar between full and reconstructed data, with only some small (mostly insignificant) differences in each. The general outline of the reconstructions' clusters are very clear and are consistent with the full data results. The boundaries of the completion clusters lose some of their smoothness, but only a few isolated pixels have been misclassified with minimal impact to the overall image.  Similarly the most important features of the absorption spectra have been preserved in the completions for DS1 and DS2, including the pre-edges, the peaks, and in particular the shifts in the absorption edges for each material. There are a few examples of noise affecting the data: DS3 is particular noisy, where for some clusters we see the sharp spikes indicative of noise. Despite this, the general outline is still consistent.

Overall, the similarity of the clusters ensures the sparse data would be interpreted in the same way as the original data, and the absorption spectra are sufficiently clear to be able to identify the materials within each cluster.

\section{Practical sparse scanning experiments}\label{sec:sparse scans}

\begin{table}[t]
    \centering
    \caption{Time and efficiency (ratio of sparse and full scan times) of implementing sparse scans during experiments.}
    \label{table:sparse times}
    \begin{tabular}{ c||c|c||c|c}\hline
        $p$ & \makecell{$T\ (hrs:min:s)$ \\ - DS4} & \makecell{$T\ (hrs:min:s)$ \\ - DS5} & \makecell{Efficiency \\ - DS4} &  \makecell{Efficiency \\ - DS5} \\
        \hline
        \hline
        1.00 & 1:22:05.5 & 4:10:36.7 & 1.00 & 1.00  \\
        \hline
        0.35 & 0:38:57.6 & 1:39:31.6 & 0.47 & 0.40    \\
        \hline
        0.25 & 0:33:29.8 & 1:21:33.8 & 0.41 & 0.33  \\
        \hline
        0.15 & 0:26:02.3 & 0:50:37.9 & 0.32 & 0.20  \\
        \hline
    \end{tabular}
\end{table}

We now bring all of the above material together to implement sparse scanning in practice. Instead of numerically sampling a full dataset, we physically implemented a sparse robust raster sampling pattern. This was done on the same specimen, with the same pixel sizes and dwell times for $p \in \{0.15,\ 0.25,\ 0.35,\ 1\} $. 

Unfortunately, it is not possible to numerically compare the results of the data due to experimental variations between the known entries in each dataset. Small changes in the ambient temperature can cause the specimen to drift very small distances between each spatial scan during the experiment. Since measurements are being taken on the micro/nano-scopic scale, this effect can create pixel level change from start to end. This is usually compensated for during the stacking process, but sparse scans have a much shorter experimental time, and registration to correct drifts cannot be performed in the same way as full datasets. The intensity of the incident beam will also vary with time and, although this can be normalized, some variation can still occur.  Because of these potential spatial discrepancies and changes in intensity and background noise, the same known entries across different scans will show different values. Thus, it is impossible to know whether the difference between the full and reconstructed data is because of differences in the measurements or due to the limitations of the completion algorithm.

\begin{figure}[H]
    \begin{subfigure}{0.24\textwidth}
        \centering
        \includegraphics[width=0.85\textwidth]{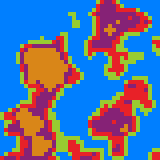}
        \caption{p = 0.15}
        \label{fig:small_015_img}
    \end{subfigure}
    \begin{subfigure}{0.24\textwidth}
        \centering
        \includegraphics[width=0.85\textwidth]{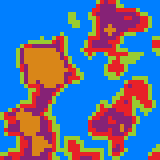}
        \caption{p = 0.25}
        \label{fig:small_025_large}
    \end{subfigure}
    \begin{subfigure}{0.24\textwidth}
        \centering
        \includegraphics[width=0.85\textwidth]{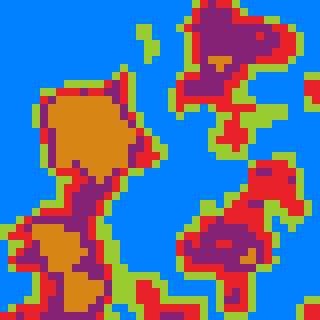}
        \caption{p = 0.35}
        \label{fig:small_035_img}
    \end{subfigure}
    \begin{subfigure}{0.24\textwidth}
        \centering
        \includegraphics[width=0.85\textwidth]{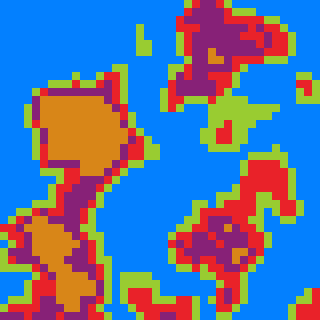}
        \caption{p = 1.00}
        \label{fig:small_100_img}
    \end{subfigure}
    
    \begin{subfigure}{0.24\textwidth}
        \centering
        \includegraphics[width=\textwidth]{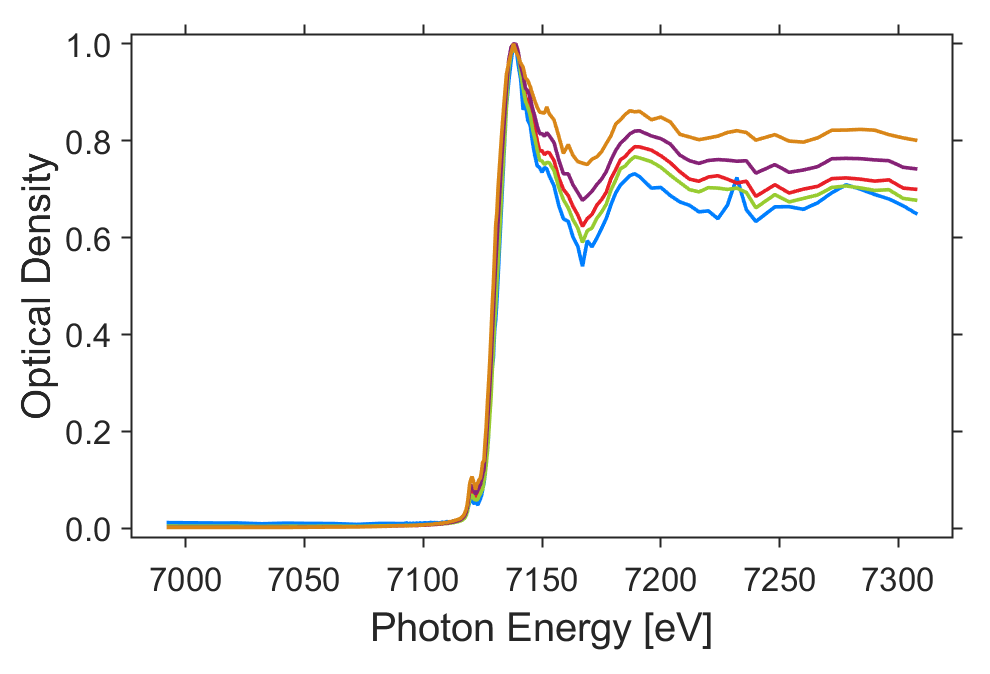}
        \label{fig:small_015_spec}
    \end{subfigure}
    \begin{subfigure}{0.24\textwidth}
        \centering
        \includegraphics[width=\textwidth]{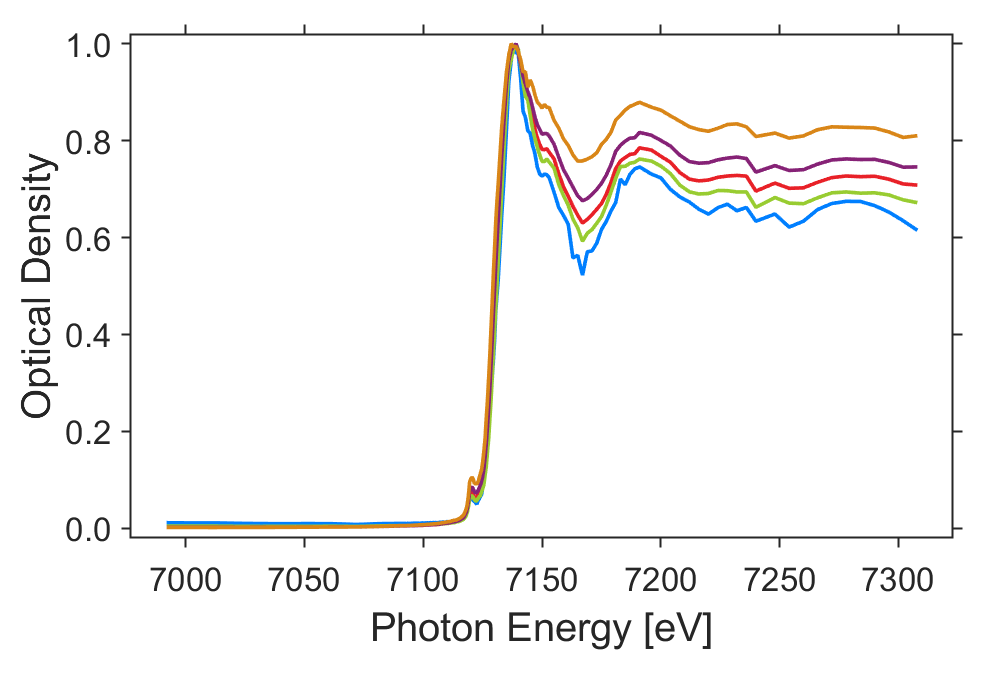}
        \label{fig:small_025_spec}
    \end{subfigure}
    \begin{subfigure}{0.24\textwidth}
        \centering
        \includegraphics[width=\textwidth]{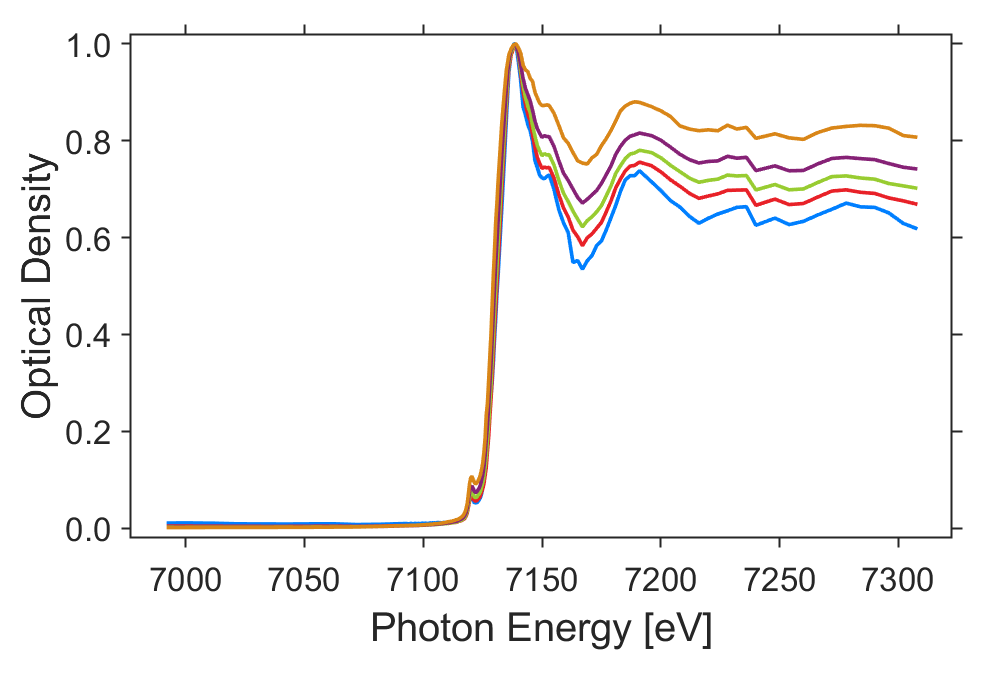}
        \label{fig:small_035_spec}
    \end{subfigure}
    \begin{subfigure}{0.24\textwidth}
        \centering
        \includegraphics[width=\textwidth]{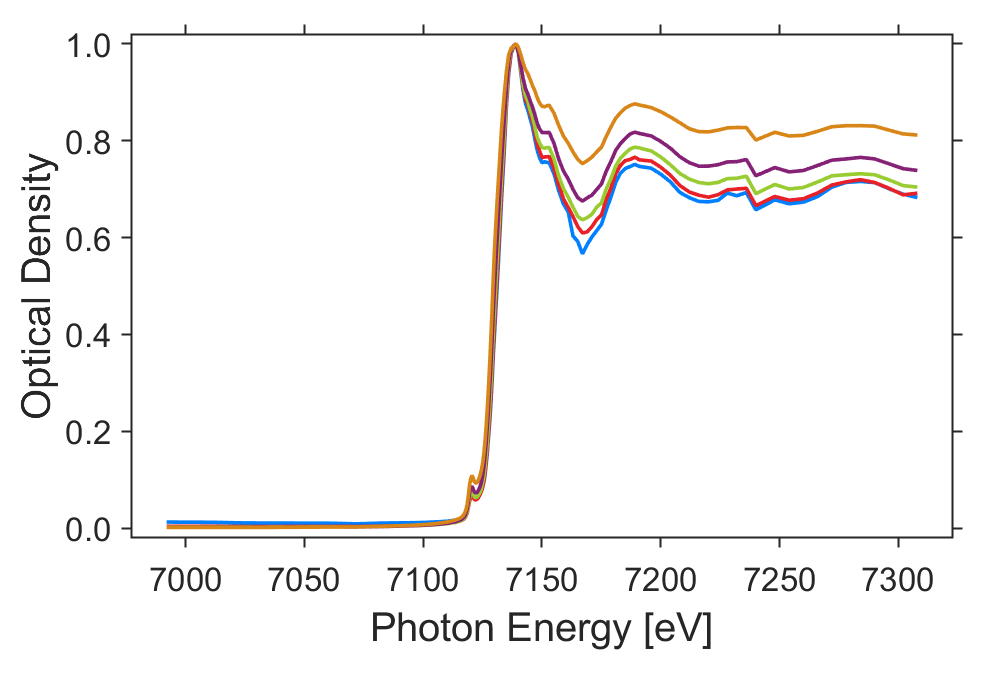}
        \label{fig:small_100_spec}
    \end{subfigure}
    \vspace{-0.5cm}
    \caption{Cluster and spectral results of sparse scanning for DS4. Measurements were taken at 15\%, 25\%, 35\%, and 100\% respectively. Cluster results computed using Mantis-xray\cite{mantis}}
    \label{fig:sparse results small}
    \vspace{0.5cm}
    
    \begin{subfigure}{0.24\textwidth}
        \centering
        \includegraphics[width=0.85\textwidth]{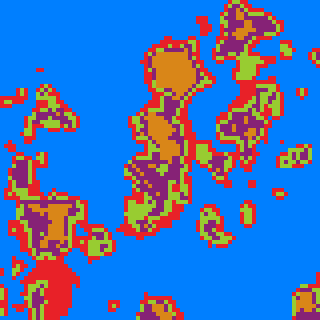}
        \caption{p = 0.15}
        \label{fig:large_015_img}
    \end{subfigure}
    \begin{subfigure}{0.24\textwidth}
        \centering
        \includegraphics[width=0.85\textwidth]{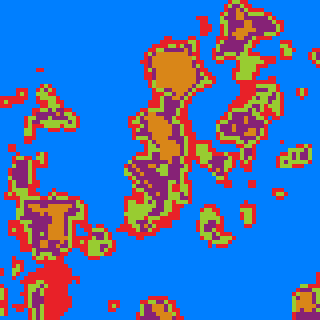}
        \caption{p = 0.25}
        \label{fig:large_025_large}
    \end{subfigure}
    \begin{subfigure}{0.24\textwidth}
        \centering
        \includegraphics[width=0.85\textwidth]{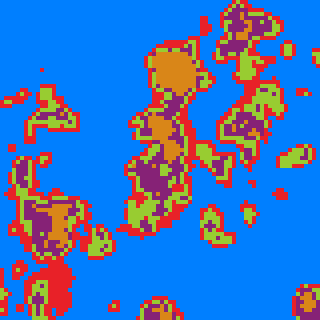}
        \caption{p = 0.35}
        \label{fig:large_035_img}
    \end{subfigure}
    \begin{subfigure}{0.24\textwidth}
        \centering
        \includegraphics[width=0.85\textwidth]{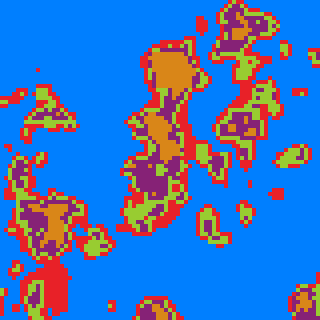}
        \caption{p = 1.00}
        \label{fig:large_100_img}
    \end{subfigure}
    
    \begin{subfigure}{0.24\textwidth}
        \centering
        \includegraphics[width=\textwidth]{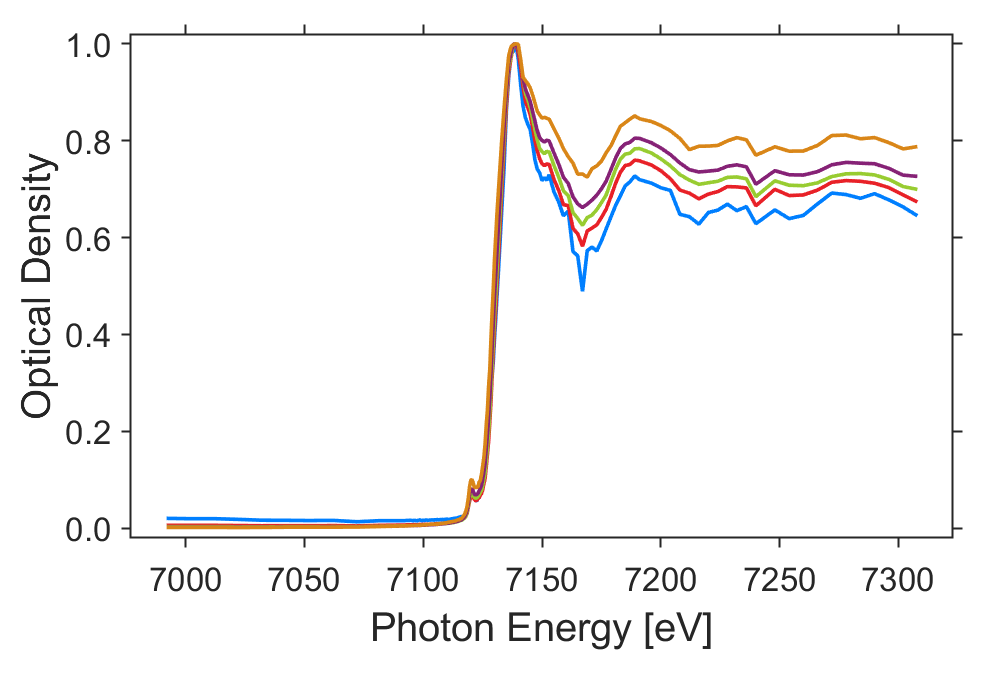}
        \label{fig:large_015_spec}
    \end{subfigure}
    \begin{subfigure}{0.24\textwidth}
        \centering
        \includegraphics[width=\textwidth]{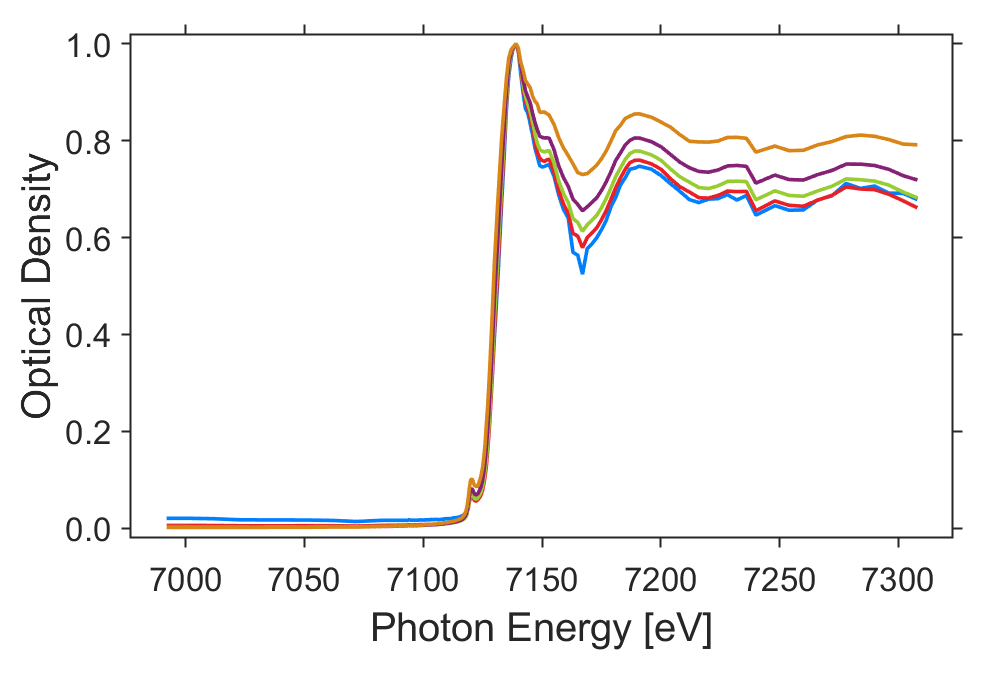}
        \label{fig:large_025_spec}
    \end{subfigure}
    \begin{subfigure}{0.24\textwidth}
        \centering
        \includegraphics[width=\textwidth]{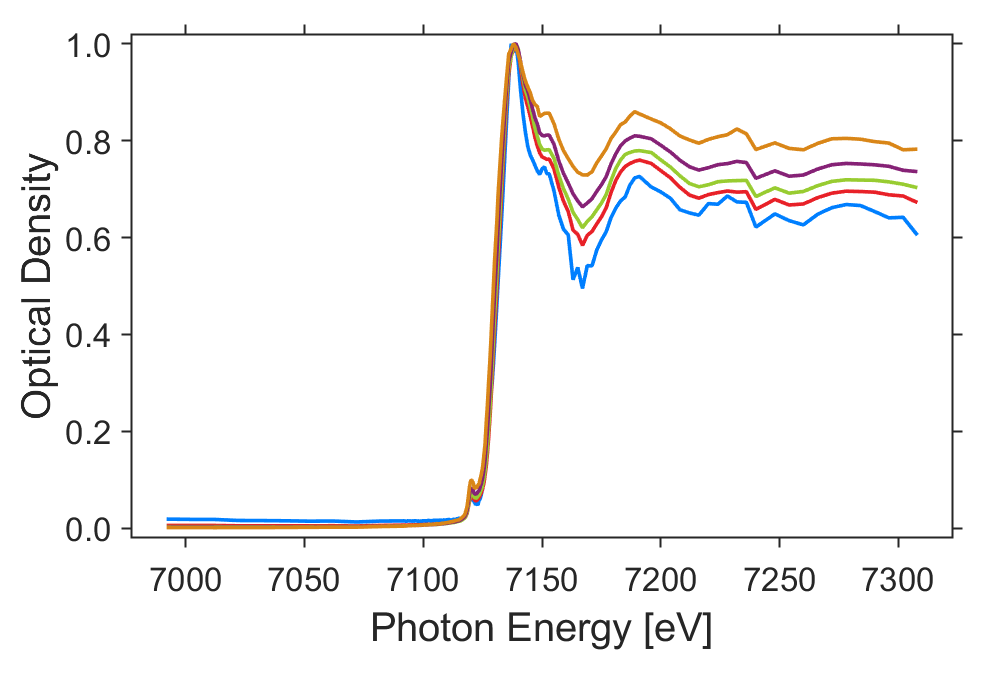}
        \label{fig:large_035_spec}
    \end{subfigure}
    \begin{subfigure}{0.24\textwidth}
        \centering
        \includegraphics[width=\textwidth]{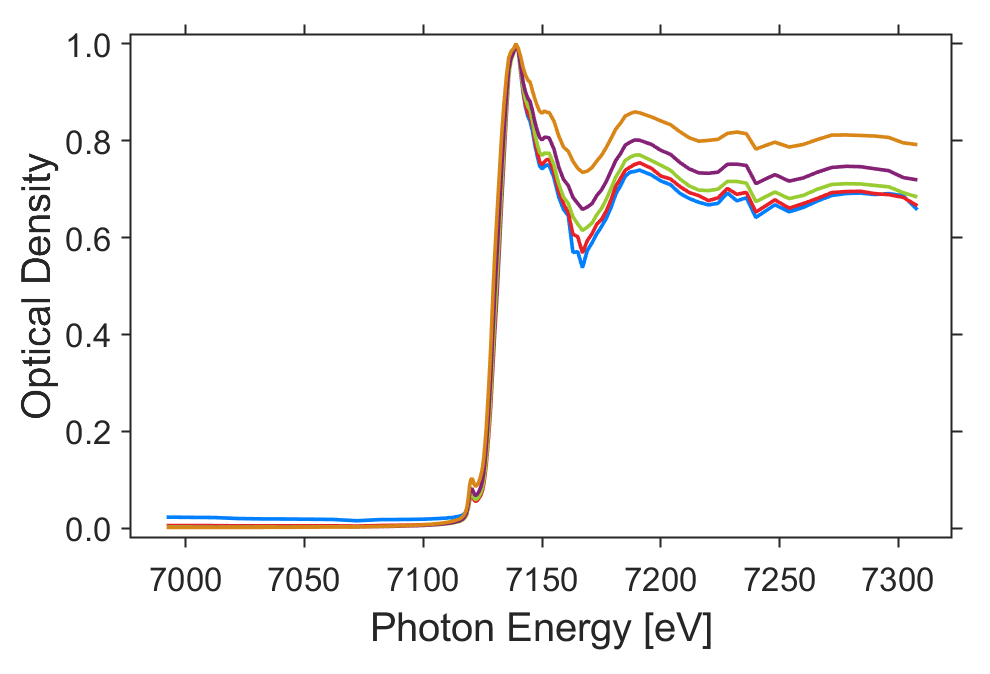}
        \label{fig:large_100_spec}
    \end{subfigure}
    \caption{Cluster and spectral results of sparse scanning for DS5. Measurements were taken at 15\%, 25\%, 35\%, and 100\% respectively. Cluster results computed by Mantis-xray \cite{mantis}.}
    \label{fig:sparse results large}
\end{figure}

Despite this, we can still \textit{visually} compare the results. For each of our sparse scans (DS4 and DS5) and at each undersampling ratio, we use LoopedASD to reconstruct the data. We equip the algorithm with the optimal completion rank for the data set, described in the supplementary documents. We use Mantis X-ray to perform the cluster analysis, setting $L = r^*$ and using 5 clusters. Recall that RTE was \textbf{not} used for these results. In Figures \ref{fig:sparse results small} and \ref{fig:sparse results large}, we plot these results. 

Once again, despite a slight loss of smoothness around cluster boundaries and some individual miss-classified pixels, the overall structure of the cluster maps is near identical and clearly shows the same variations across the scans. Similarly, there are a few sharp discontinuities in the reconstructed absorption spectra, especially for lower undersample ratios, but each materials' XANES are still clearly identifiable. 

After extracting the timestamps from the files' meta data, we can compare the run time for each sparse experiment, which have been summarised in Table~\ref{table:sparse times}.

The true time efficiency will never exactly match the undersampling ratio used due to `dead time' as the scanner resets and for critical machine processes. Despite this, we see huge gains for the lowest undersample ratios, and a significant increase in experimental efficiency. In particular, we see greater improvements for the larger region of DS5, due to the higher measurement to dead time ratio. Another benefit for scanning larger regions is that the completion algorithms become more efficient. Because it is approximately low rank, the total number of entries grows much faster than the number of degrees of freedom - i.e. larger datasets are recoverable at lower undersampling ratios. Thus, it is in the interest of researchers using this approach to scan larger areas to improve both the scanning efficiency, and the quality of reconstruction.

\section{Conclusion}

We have demonstrated a new undersampling approach for spectromicroscopy data acquisition. By taking advantage of the inherent low rank structure of the datasets, we have produced algorithms that can accurately recover unknown entries from as little as $15\%$ of the measurements when raster sampling. We have illustrated the robustness of these algorithms to machine artefacts, and how to derive parameters like the completion rank from the sparse data itself.

Finally, we showed the minimal impact reconstructing sparse data has on the cluster analysis that is currently used to interpret the spectromicroscopy data. By implementing these methods, we can conduct experiments at a much faster rate, over larger areas, and with lower dose on the sample: the method consistently produces near-identical results using $20\%$ of the measurements, with potential improvements to 15\% for larger samples. Alternate sampling schemes, compatible with continuous motion scans, may be able to produce further reductions. These improvements will help in the development of in-situ spectro-microscopy measurements and future developments will reduce times in areas such as XANES nano-tomography and nano-EXAFS, which are currently limited in their application by long acquisition times. 

\section{Backmatter}

\begin{backmatter}
\bmsection{Funding}
This research is partially supported by a University of Bath scholarship via the EPSRC Centre for Doctoral Training in Statistical Applied Mathematics at Bath (SAMBa) under project EP/S022945/1.

This research is partially supported by Diamond Light Source Plc.

Content in the funding section will be generated entirely from details submitted to Prism. Authors may add placeholder text in the manuscript to assess length, but any text added to this section in the manuscript will be replaced during production and will display official funder names along with any grant numbers provided. If additional details about a funder are required, they may be added to the Acknowledgments, even if this duplicates information in the funding section. See the example below in Acknowledgements.

\bmsection{Acknowledgments}
We acknowledge Diamond Light Source for time on Beamline/Lab I14 under Proposal MG31039.

The authors gratefully acknowledge the University of Bath’s Research Computing Group \\ (doi.org/10.15125/b6cd-s854) for their support in this work, see \cite{bath_computing}.

\bmsection{Disclosures}
The authors declare no conflicts of interest.

\bmsection{Data availability} Data underlying the results presented in this paper are not publicly available at this time but may be obtained from the authors upon reasonable request.

\end{backmatter}

\bibliography{sample}

\newpage

\section{Supplemental}

\subsection{PCA and RTE}
We begin by describing the Principal Component Analysis (PCA) and cluster analysis of spectromicroscopy data in more detail. Additionally, we explain how Reduced Thickness Effects (RTE) are implemented. This is based on the methods outlined in \cite{Cluster_Analysis}.

Given our flattened x-ray spectromicroscopy dataset $A\in \mathbb{R}^{n_E\times N}$, we wish to separate the data into its principal components by imposing the following decomposition,
\begin{equation}\label{eq:CR}
    A = CR,
\end{equation}
with $C\in \mathbb{R}^{n_E \times n_E}$ and $R \in \mathbb{R}^{n_E \times N}$. This can be done either by computing the covariance matrix of $A$, or using the SVD (singular value decomposition). Both methods are equivalent; we shall describe the covariance approach here.

First, we compute the spectral covariance matrix of $A$, $B \in \mathbb{R}^{n_E \times n_E}$, then perform an eigen-decomposition on $B$ to get,
\begin{align}
    B &= AA^{T},\\
    BC&=C\Lambda,
\end{align}
where $\Lambda \in \mathbb{R}^{n_E \times n_E}$ is a diagonal matrix of the eigenvalues of $B$, and the columns of $C \in \mathbb{R}^{n_E \times n_E}$ are the corresponding eigenvectors. Since the columns of $C$ are orthogonal, its left inverse is equal to $C^{T}$. Note that we compute the spectral covariance since typically $n_E << N$.

To complete the $C$-$R$ decomposition of $A$ in Eq. (\ref{eq:CR}), we compute 
\begin{equation}
    R = C^{-1} A = C^TA,
\end{equation}
The columns of $C$ and rows of $R$ are the principal components of $A$, and the corresponding eigenvalues in $\Lambda$ describe their significance in the dataset $A$. Additionally, the columns of $C$ can be thought of as abstract absorption spectra and the rows of $R$ as abstract spatial maps - linear combinations of true spectra and spatial maps, without any physical interpretation.

To compute the PCA's low rank approximation, we simply take the first $L$ columns of $C$ and the first $L$ rows of $R$, and discard the remaining data. Thus we get,
\begin{equation}
    A' = C' R',
\end{equation}
where $A' \in \mathbb{R}^{n_E \times N}$, $C' =  C(:\ ,\ 1:L) \in \mathbb{R}^{n_E \times L}$ and $R' = R(1:L\ ,\ :) \in \mathbb{R}^{L \times N}$. Note, we have used MATLAB-like notation to partition the rows and columns of the matrices. To maximise the variation of the approximation while minimising the rank, $L$ is chosen to be at the elbow (point of maximum curvature) of the eigenvalues, $\Lambda$. Again, this is equivalent to setting $L$ to be the elbow of the singular values of $A$.

\begin{figure}
    \begin{subfigure}{0.33\textwidth}
        \centering
        \includegraphics[width = 0.95\textwidth]{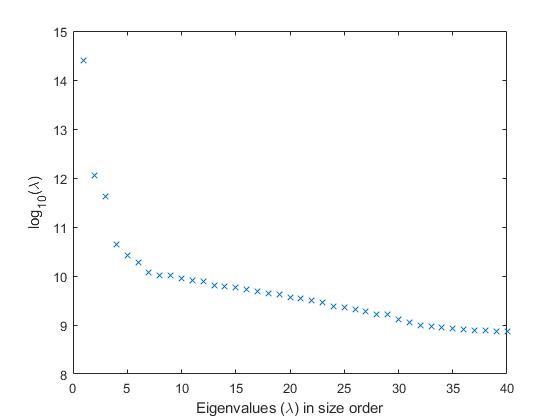}
        \caption{Eigenvalues (diag. entries of $\Lambda$)}
        \label{fig:evalues}
    \end{subfigure}
    \begin{subfigure}{0.35\textwidth}
        \centering
        \includegraphics[width = 0.9\textwidth]{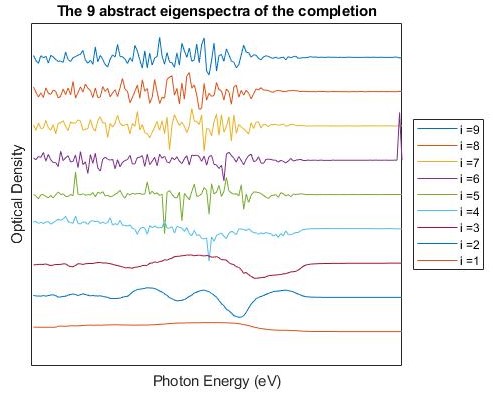}
        \caption{Eigenspectra (columns of $C$)}
        \label{fig:espectra}
    \end{subfigure}
    \begin{subfigure}{0.3\textwidth}
        \centering
        \includegraphics[width = 0.9\textwidth]{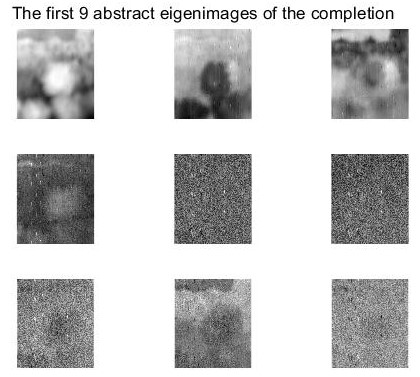}
        \caption{Eigenimages (rows of $R$)}
        \label{fig:eimages}
    \end{subfigure}    
    \caption{PCA results of DS2. KNEEDLE identified the elbow of the eigen values at $4$. Notice the first 4 eigenspectra and eigenimage contain useful information, after which we see the components are corrupted by excessive noise.}
    \label{fig:PCAonDS2}
\end{figure}

In Figure \ref{fig:PCAonDS2} we illustrate the eigenvalues of DS2, alongside the first 9 eigenspectra and eigenimages (columns of $C$ and rows of $R$ respectively). KNEEDLE, \cite{kneedle}, was used and identified the elbow as $L = 4$. In Figures \ref{fig:espectra} \& \ref{fig:eimages}, we see that $L = 4$ correctly determines the number of components that show significant variation and little noise: the first four spectra are much more smooth than the rest, and all eigenimages after the $4^{th}$ show the `salt and pepper' pattern that is indicative of noise. In practice we use $L=5$ to ensure we capture all the variation in the data.

To identify the different materials in the specimen, we now use cluster analysis on the columns of $R'$ to collect together pixels with similar combinations of abstract absorption spectra. The number of cluster centres must be set, and should be greater than, or equal to, the number of materials within the specimen. Several different cluster algorithms may be used, but we generally use a standard implementation of kmeans. Averaging the x-ray absorption spectrum over all the pixels in each cluster provides the corresponding absorption spectra of that material, with a much higher signal to noise ratio.

As was explained in the main text, it can be seen that the first component is simply averaged over all pixels and is associated more with the overall thickness of the specimen \cite{Cluster_Analysis}. Occasionally, we wish to avoid the thickness of the specimen skewing the clustering results by reducing the effect of the first principal component. To implement reduced thickness effect (RTE), we simply ignore the most significant component and set $C' =  C(:\ ,\ 2:L)$ and $R' = R(2:L\ ,\ :)$ before applying the cluster analysis as usual. Note that other approaches are available, such as scaling the principal components.

\subsection{Robust raster sampling}

Given a dense dataset $A$, we model the sparse scan $\mathcal{P}_\Omega (A)$ by computing the following projection,
\begin{equation}
    \mathcal{P}_\Omega (A) = \Omega \circ A,
\end{equation}
where $\circ$ is the Hadamard product, and $\Omega \in \{0,1\}^{n_E \times N}$ is the sampling pattern indicating the location of the known entries. We now discuss different methods of producing the randomised sampling patterns to be used in sparse scans. 

The most common sampling model used is Bernoulli sampling, where each data-point is sampled i.i.d (independent and identically distributed) with probability $p$. To improve the efficiency of sparse spectromicroscopy, we instead use Raster sampling, which scans physical rows of the specimen together i.i.d.  with probability $p$. This way, rows that aren't sampled can be passed over quickly.

One disadvantage is that, for low undersampling ratios, the probability that a row is not sampled at all increases sharply. 

It is clear that it is impossible to recover pixels that are not sampled for any energy level using Low Rank Matrix Completion. In fact, it seems the more frequently a point is sampled, the easier it is to recover the surrounding missing entries (e.g. one pixel at different energy levels, or different pixels at one energy level). To ensure all rows of the specimen are scanned, and to promote a more even sampling pattern, we use a robust variation of Raster Sampling called \emph{Robust Raster Sampling}.

Robust Raster Sampling can be implemented either by varying the probability of each row, or by discretely restricting which rows can be scanned. Here we describe the discrete method.

Begin with a set $v = [n_1] = \{1,\ ...\ , n_1 \}$ that indexes the rows that have yet to be scanned (recall that $n_1$ is the number of rows on the sample). Next, compute the expected number of rows to be sampled at each energy, $p n_1$. For each energy level, sample $pn_1$ rows uniformly at random from $v$, removing the sampled rows from $v$. If $pn_1$ is not an integer, simply take the ceiling or floor at ranodm using the decimal as the probability. Repeat this process for each energy level $E_i$, $i = 1, ... , n_E$. For energy levels where $pn_1 > |v|$ (i.e. there are fewer than $pn_1$ rows remaining in $v$), we must first record the deficit $m = pn_1 - |v|$ before sampling the last rows in $v$. We now repopulate $v$, withholding only the rows that have already been sampled on this energy level (to avoid sampling them twice). Finally, we sample the remaining $m$ rows from the repopulated set $v$ before moving to the next energy level.

In short, Robust Raster sampling ensures that every row is sampled once before any can be sampled twice. Additionally, be ensuring $pn_1$ rows are sampled for each energy level, the known entries are more evenly distributed across $\mathcal{P}_\Omega(A)$.

\subsection{Details on ASD}

To recover the missing entries of $\mathcal{P}_{\Omega}(A)$, ASD \cite{ASD} imposes the decomposition $A = XY^T, X \in \mathbb{R}^{n_E \times r}, Y \in \mathbb{R}^{N \times r}$ and seeks to minimise the objective function:
\begin{equation}
    f(X,Y) = \frac{1}{2}||\mathcal{P}_{\Omega}(A - XY^T)||^2_F.
\end{equation}
This is done by alternately fixing one component and minimising the other using gradient descent with exact step sizes. For this simple procedure, the directions $d_X,\ d_Y$ and step sizes $t_X,\ t_Y$ are computed to give the next iterates:  
\begin{equation}
    X_{i+1} = X_{i} + d_X t_X \qquad \text{and} \qquad Y_{i+1} = Y_i + d_Y t_Y.
\end{equation}
We write $f(X,Y)$ as $f_Y(X)$ when $Y$ is fixed, and $f_X(Y)$ when $X$ is fixed. Thus, we have gradients,
\begin{equation}
    \nabla f_Y(X) = -(\mathcal{P}_{\Omega}(A) - \mathcal{P}_{\Omega}(XY^T))Y \qquad \text{and} \qquad \nabla f_X(Y) = -X^T(\mathcal{P}_{\Omega}(A) - \mathcal{P}_{\Omega}(XY^T)).
\end{equation}
For gradient descent, we use the step directions, 
\begin{equation}
    d_X = - \nabla f_Y(X) \qquad \text{and} \qquad d_Y = - \nabla f_X(Y).
\end{equation}
Finally, we can compute the exact step sizes required for steepest descent. This is done by minimising,
\begin{align}
    g_X(t) &= f_Y(X + td_X), & \text{and} \quad g_Y(t) &= f_X(Y + td_Y),\\
    &= \frac12||\mathcal{P}_{\Omega}(A) - \mathcal{P}_{\Omega}((X + td_X)Y^T)||^2_F, & &= \frac12||\mathcal{P}_{\Omega}(A) - \mathcal{P}_{\Omega}(X(Y + td_Y)^T)||^2_F .
\end{align}
Indeed, we can compute the step sizes exactly, giving
\begin{align}
    t_X  &= \argmin{g_X(t)} & \text{and}\qquad \qquad t_Y &= \argmin{g_Y(t)} \\
    &= \frac{||\nabla f_Y(X)||^2_F}{||\mathcal{P}_{\Omega}(\nabla f_Y(X)Y^T)||^2_F} & &= \frac{||\nabla f_X(Y)||^2_F}{||\mathcal{P}_{\Omega}(X[\nabla f_X(Y)]^T)||^2_F}
\end{align}

It can be seen that, following the updates to $X_{i+1}$, $Y_{i+1}$, the residuals can be written as
\begin{align}
    \text{res}_{X_{i+1}} &= \mathcal{P}_{\Omega}(A- X_{i+1}Y_i^T)                   & \text{res}_{Y_{i+1}} &= \mathcal{P}_{\Omega}(A- X_{i+1}Y_{i+1}^T) \\
    &= \mathcal{P}_{\Omega}(A- (X_i + t_{X_i}d_{X_i})Y_i^T)                         & &= \mathcal{P}_{\Omega}(A- X_{i+1}(Y_i + t_{Y_i}d_{Y_i})^T)\\
    &= \mathcal{P}_{\Omega}(A- X_iY_i^T) - t_{X_i}\mathcal{P}_{\Omega}(d_{X_i}Y_i^T)& &= \mathcal{P}_{\Omega}(A- X_{i+1}Y_i^T) - t_{Y_i}\mathcal{P}_{\Omega}(X_id_{Y_i}^T)\\
    &= \text{res}_{Y_i} - t_{X_i}\mathcal{P}_{\Omega}(d_{X_i}Y_i^T)                 & &= \text{res}_{Y_i} - t_{Y_i}\mathcal{P}_{\Omega}(X_id_{Y_i}^T)
\end{align}

This allows for an efficient implementation. Since $\mathcal{P}_{\Omega}(d_{X_i}Y_i^T)$ is already formed when computing $t_{X_i}$, we can avoid the matrix-matrix multiplication required to calculate the residual, and simply update it for each iteration. Using this update, the per iteration cost of this algorithm is $8\Omega r$ \cite{ASD}, making it very efficient. The method is summarised in Algorithm \ref{Alg:ASD}.

\begin{algorithm}
    \caption{Alternating Steepest Decent (ASD)}
    \label{Alg:ASD}
    \textbf{Input}: $\mathcal{P}_{\Omega}(A),\ X_0 \in \mathbb{R}^{n_E \times r},\ Y_0 \in \mathbb{R}^{r \times N},\ kmax$
    \begin{algorithmic}
    \For{$i=1$ \textbf{to} kmax}
    \State $\hphantom{\nabla f_{X_{i+1}}(Y_i)}\mathllap{\nabla f_{Y_i}(X_i)}  = -res_{Y_i}Y_i^T$ 
    \State $\hphantom{\nabla f_{X_{i+1}}(Y_i)}\mathllap{d_{X_i}} = -\nabla f_{Y_i}(X_i)$
    \State $\hphantom{\nabla f_{X_{i+1}}(Y_i)}\mathllap{t_{X_i}}  = ||\nabla f_{Y_i}(X_i)||^2_F\ /\ ||P_{\Omega}(\nabla f_{Y_i}(X_i)Y_i)||^2_F$
    \State $\hphantom{\nabla f_{X_{i+1}}(Y_i)}\mathllap{X_{i+1}} = X_i + t_{X_i} d_{X_i}$
    \State $\hphantom{\nabla f_{X_{i+1}}(Y_i)}\mathllap{\text{res}_{X_{i+1}}} =  \text{res}_{Y_i} - t_{X_i}\mathcal{P}_{\Omega}(\nabla f_{Y_i}(X_i)Y_i^T)$
    \ \\
    \State $\nabla f_{X_{i+1}}(Y_i) = -X^T_{i+1} (P_{\Omega}(A) - P_{\Omega}(X_{i+1} Y_i))$
    \State $\hphantom{\nabla f_{X_{i+1}}(Y_i)}\mathllap{d_{Y_i}} = -\nabla f_{X_{i+1}}(Y_i)$
    \State $\hphantom{\nabla f_{X_{i+1}}(Y_i)}\mathllap{t_{Y_i}}  = ||\nabla f_{X_{i+1}}(Y_i)||^2_F\ /\ ||P_{\Omega}(X_{i+1}\nabla f_{X_{i+1}}(Y_i))||^2_F$
    \State $\hphantom{\nabla f_{X_{i+1}}(Y_i)}\mathllap{Y_{i+1}} = Y_i + t_{Y_i} d_{Y_i}$
    \State $\hphantom{\nabla f_{X_{i+1}}(Y_i)}\mathllap{\text{res}_{Y_{i+1}}} =  \text{res}_{X_i} - t_{Y_i}\mathcal{P}_{\Omega}(X_i^T\nabla f_{X_i}(Y_i))$
    \   \\  
    \If{\text{Stopping Conditions Reached}}
    \State \textbf{break}
    \EndIf
    \EndFor
    \end{algorithmic}
\end{algorithm}

\subsection{Choosing the completion rank}\label{sec: choose rank}

It has been noted before that the completion rank of the dataset must be set before implementing ASD or LoopedASD. In Section 2 of the main document, we derived the low rank nature of spectromicroscopy datasets, but correctly identifying the optimal rank will determine the success of the completion algorithm used. We have noted in Section 4 of the main document that datasets with lower approximate ranks require fewer known entries to produce good reconstructions, making the sparse experiments more efficient. It is therefore crucial that robust and reliable methods for estimating the rank of a sampled dataset are developed. By accurately estimating this value, the reconstruction should capture the full variation of the data, while filtering out as much of the noise as possible; this will ensure the sparse clustering results are similar to the full-data case. 

The first step is to identify the most appropriate completion rank using the full datasets. Recall the Eckart-Young-Mirsky (EYM) Theorem \cite{E-Y-M_Theorem}, which provides the \textbf{minimum approximation error} in the frobenius norm for a rank-$k$ approximation of $A$, denoted $A^{(k)}$. It should be noted that this approach cannot be used on sparse data. 

\begin{figure}
    \begin{subfigure}{0.33\textwidth}
        \centering
        \includegraphics[width = \textwidth]{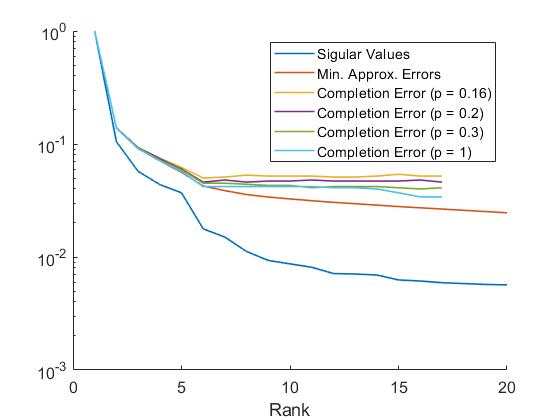}
        \caption{DS1}
        \label{fig:SV DS1}
    \end{subfigure}
    \begin{subfigure}{0.33\textwidth}
        \centering
        \includegraphics[width = \textwidth]{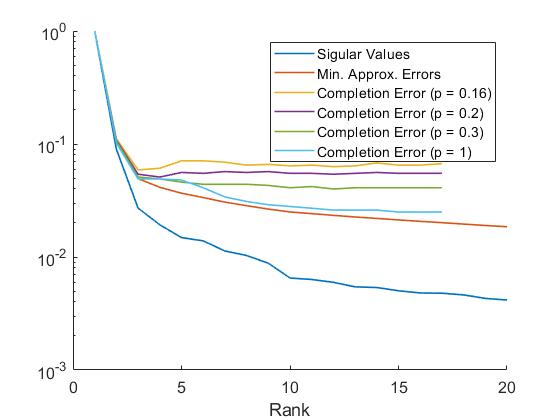}
        \caption{DS3}
        \label{fig:SV DS3}
    \end{subfigure}
    \begin{subfigure}{0.33\textwidth}
        \centering
        \includegraphics[width = \textwidth]{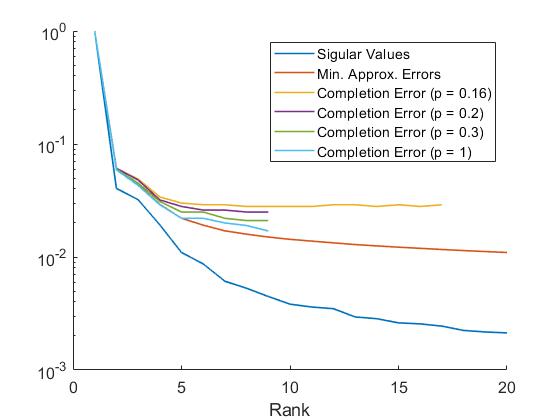}
        \caption{DS5}
        \label{fig:SV DS5}
    \end{subfigure}
    \caption{Plots illustrating the Singular Values, the minimum approximation errors and mean ASD completion errors for each rank. Note the singular values and minimum approximation errors have been normalised to enable a qualitative comparison. The minimum approximation errors are computed using the Eckart-Young-Mirsky Theorem (Eq. \ref{eq:EYM Thm}). Completion errors, $e_c$, are computed as the relative Frobenius norm of the difference between the full data and the reconstruction (see Eq(16) in the main text). The values illustrated are means taking over $16$ independent completions. We see the similar patterns indicate the optimal completion rank is the same as the optimal approximation rank for standard spectromicroscopy PCA.}
    \label{fig:min approx error}
\end{figure}

By repeatedly sampling and completing full datasets for ranks $r = 1,...,25$ and a range of undersampling ratios, we can determine how the  \textbf{completion results} vary with the completion rank. In Figure \ref{fig:min approx error}, we have plotted the singular values (dark blue), the minimum completion errors (orange), and the completion results for each independent dataset available (DS1, DS3, DS5).

We can see that the curve of the minimum approximation errors produced by the EYM Theroem have similar curved shapes to the plot of SVs. This similarity is due to the dominance of the larger SVs in the estimate. When the SVs decay rapidly, $\sigma_{k+1}^2 >> \sigma_{i}^2$ for $i= (k+2)\ ,\ ...\ ,\ n$; thus $||A - A^{(k)}||^2 \approx \sigma_{k+1}$ and also decays rapidly. Then, as the SV plot flattens, we have that $\sigma_{k+1} \approx \sigma_{k+2} \approx \sigma_{k+3} \approx ...$ and the difference between consecutive minimum approximation errors decreases, flattening this plot as well (orange plot in Figure \ref{fig:min approx error}).

Naturally, the minimum approximation error is a lower bound for any completion algorithm. As can be seen in Figure \ref{fig:min approx error}, the errors will easily achieve their optimal values when using lower ranks. For higher completion ranks, the flattening of the EYM plot means the completion errors must flatten as well. Despite the fact that the bound decrease slowly, it becomes increasingly difficult for algorithms to produce minimal errors due to slower convergence rates, so the flattening of the observed completion errors is more severe. Overall, we see that the average completion errors plots a similar curve to the SVs and minimum approximation errors - initially decreasing rapidly before flattening and remaining relatively constant. 

Consider now what are the properties of the optimal rank: it should minimise the completion error using the smallest completion rank. This is precisely the same problem as selecting the rank $L$ for the PCA in the main document, and given the connection we have established between the SV plot and the completion error plot, we apply the same selection method here as well.
Once again, we seek the elbow point (the point of maximum curvature) of the known completion errors, which we label $r^*$. We see in Figure \ref{fig:min approx error} that for $r<r^*$, the both the average and minimum completion errors decrease rapidly, thus increasing the rank will significantly improve the reconstruction; for $r>r^*$, there is little difference in the mean completion error, and increasing the rank will not yield better results. Thus, $r^*$ reaches the optimal balance - it allows ASD to capture almost all the variation in the data, while remaining low to improve the probability of completion, especially  with lower undersample ratios.

Now that we have a target, we can develop methods for identifying $r^*$ from the sparse samples, $\mathcal{P}_{\Omega}(A)$. The general outline will be to compute short, approximate completions for each trial rank, compute and plot the residual-norm, then find the elbow point using KNEDDLE. Since the data is sparse, we can only evaluate a completion using the residual-norm and not the full completion error, i.e. we must compare back to the known entries used for the completion. To avoid overfitting, we take a cross validation approach.

Iterating over the trial completion rank, $r$, we partition the known entries into $k$ subsets and select one at random to be the validation set - the remaining entries become the training data. From experience, it is very unlikely a dataset will have an approximate rank higher that 15, so iterating through 20 trial ranks is sufficient in general. This value is easily increased if needs be. 

For each trial rank $r = 1,...,20$, the training data is completed using ASD, and the validation error is computed by taking the relative Frobenius norm over the validation set. We set ASD to run for $n\_{it}$ iterations; this parameter, alongside the number of partitions $k$, can be tuned to adjust the efficiency and accuracy of this process. Once all validation errors have been computed, we use KNEEDLE to compute the elbow.

This new procedure for identifying the completion rank is summarised below:

\begin{equation}
    \begin{cases}
        \textbf{for}\ r=1,...,20:\\
        \quad \text{Partition the known entries to get } \{B_1, ... B_k\}, \text{ so that } \Omega = \bigcup_{j = 1}^kB_k\ . \\
        \quad \text{Choose } i \in [k] \text{ uniformly at random.}\\
        \quad \text{Set } \hat{B} = B_i \text{ to be the validation set and } \Omega \backslash B_i \text{ to be the training set.}\\
        \quad \text{Complete the training data using ASD with completion rank } r \text{ and } n\_{it} \text{ iterations.}\\
        \quad \text{Compute the validation error} \quad \frac{||\mathcal{P}_{\hat{B}}(A) - \mathcal{P}_{\hat{B}}(A^*) ||_F^2}{||\mathcal{P}_{\hat{B}}(A)||_F^2}\text{, where } A^* = (X^*)^{(r)} (Y^*)^{(r)}.\\
        \quad \text{Record the validation error as the } r^{th} \text{ trial error.} \\
        \textbf{end}\\
        \text{Apply KNEEDLE to identify the elbow of the trial errors}.
    \end{cases}
\end{equation}

Following some testing of the new method, it was quickly found that there was little change in accuracy for $k>10$. For $k \leq 10$, the training sets were made too small by the partitioning to reliably reconstruct the datasets. From here on, we fix $k = 20$, and we found that $n_{it} = 300$ iterations of ASD were generally sufficient to ensure reliable results across all undersample ratios and completion ranks. These results averaged slightly below the previously computed optimal rank, with a difference of no more than 1. Since it is better to overestimate the completion rank than under estimate it, we add 2 to the result to ensure we've captured the full variation within the data.

\end{document}